\documentclass[a4paper]{aa}  

\usepackage{siunitx}
\usepackage{stfloats}

\usepackage{graphicx}
\usepackage{txfonts}
\usepackage[colorlinks=true,
            linkcolor=red,
            urlcolor=blue,
            citecolor=blue]{hyperref}

\newif\ifresub
\newif\iffinsub

\begin{document}
    
    \resubfalse
    \finsubfalse
    
    \newcommand{\rs}[1]{\ifresub\textbf{#1}\else#1\fi}
    \newcommand{\fis}[1]{\iffinsub\textbf{#1}\else#1\fi}

    \title{Large Interferometer For Exoplanets (\emph{LIFE}):}%
    \subtitle{II. Signal simulation, signal extraction and fundamental exoplanet parameters from single epoch observations}
    \titlerunning{\emph{LIFE}: II. Signal simulation, signal extraction and exoplanet parameter estimates}

    \author{
            Felix Dannert\inst{\ref{eth},\ref{nccr}} \thanks{Equal contribution, correspondence to: \href{mailto:fdannert@phys.ethz.ch}{fdannert@phys.ethz.ch}}\and
            Maurice Ottiger\inst{\ref{eth}}\thanks{Equal contribution}\and
            Sascha P. Quanz\inst{\ref{eth},\ref{nccr}}\and
            Romain Laugier\inst{\ref{leuven}}\and
            Emile Fontanet\inst{\ref{eth}}\and
            Adrian Gheorghe\inst{\ref{eth}}\and
            Olivier Absil\inst{\ref{liege}}\thanks{F.R.S.-FNRS Research Associate}\and
            Colin~Dandumont\inst{\ref{csl}}\and
            Denis Defr\`ere\inst{\ref{leuven}}\and
            Carlos Gasc\'on\inst{\ref{bella}}\and
            Adrian M. Glauser\inst{\ref{eth}}\and
            Jens Kammerer\inst{\ref{stsci}}\and
            Tim~Lichtenberg\inst{\ref{oxford}}\and
            Hendrik Linz\inst{\ref{mpia}}\and
            Jer\^ome Loicq\inst{\ref{csl},\ref{tudelft}}\and
            the \emph{LIFE} collaboration\inst{\ref{life}}
            }
    \authorrunning{Dannert et al.}
    
    \institute{
    ETH Zurich, Institute for Particle Physics \& Astrophysics, Wolfgang-Pauli-Str. 27, 8093 Zurich, Switzerland\label{eth}
    \and
    National Center of Competence in Research PlanetS (www.nccr-planets.ch)\label{nccr}
    \and 
    Institute of Astronomy, KU Leuven, Celestijnenlaan 200D, 3001, Leuven, Belgium\label{leuven}
    \and 
       STAR Institute, University of Li\`ege, 19C all\'ee du Six Ao\^ut, 4000 Li\`ege, Belgium\label{liege}
       \and
          Centre Spatial de Li\`ege, Universit\'e de Li\`ege, Avenue Pr\'e-Aily, 4031 Angleur, Belgium\label{csl}
    \and
    Institut de Ci\`encies de l'Espai (ICE, CSIC), Campus UAB, C/Can Magrans s/n, 08193 Bellaterra, Spain\label{bella}
    \and
        Space Telescope Science Institute, 3700 San Martin Drive, Baltimore, MD 21218, USA\label{stsci}
    \and
       University of Oxford, Atmospheric, Oceanic and Planetary Physics, Department of Physics, Sherrington Road, Oxford OX1 3PU, United Kingdom\label{oxford}
    \and
   Max-Planck-Institut f\"ur Astronomie, K\"onigstuhl 17, 69117 Heidelberg, Germany\label{mpia}
   \and
   Faculty of Aerospace Engineering, Delft University of Technology, Kluyverweg 1, 2629 Delft, Netherlands\label{tudelft}
   \and
    \url{www.life-space-mission.com}\label{life}
    }
    
   \date{Received: <date> / Accepted: <date>}
    
    \abstract
    {The \emph{Large Interferometer For Exoplanets (LIFE)} initiative is developing the science and a technology roadmap for an ambitious space mission featuring a space-based mid-infrared (MIR) nulling interferometer in order to detect the thermal emission of hundreds of exoplanets and characterize their atmospheres.}
    {In order to quantify the science potential of such a mission, in particular in the context of technical trade-offs, an instrument simulator is required. In addition, signal extraction algorithms are needed to verify that exoplanet properties (e.g., angular separation, spectral flux) contained in simulated exoplanet datasets can be accurately retrieved.}
    {We present \textsc{LIFEsim}, a software tool developed for simulating observations of exoplanetary systems with an MIR space-based nulling interferometer. It includes astrophysical noise sources (i.e., stellar leakage and thermal emission from local zodiacal and exo-zodiacal dust) and offers the flexibility to include instrumental noise terms in the future. \rs{Here, we provide some first quantitative limits on instrumental effects that would allow the measurements to remain in the fundamental noise limited regime.} We demonstrate updated signal extraction approaches to validate signal-to-noise ratio (SNR) estimates from the simulator. Monte-Carlo simulations are used to generate a mock  survey of nearby terrestrial exoplanets and determine to which accuracy fundamental planet properties can be retrieved.}
    {\textsc{LIFEsim} provides an accessible way for predicting the expected SNR of future observations as a function of various key instrument and target parameters. The SNRs of the extracted spectra are photon-noise dominated, as expected from our current simulations. Signals from multi-planet systems can be reliably extracted. From single epoch observations in our mock survey \rs{of small ($R < 1.5 R_\mathrm{Earth}$) planets orbiting within the habitable zones of their stars}, we find that typical uncertainties in the estimated effective temperature of the exoplanets are $\lesssim$10\%, for the exoplanet radius $\lesssim$20\%, and for the separation from the host star $\lesssim$2\%.
    SNR values obtained in the signal extraction process deviate less than 10\% from purely photon-counting statistics based SNRs.}
    {\textsc{LIFEsim} has been sufficiently well validated so that it can be shared with a broader community interested in quantifying various exoplanet science cases that a future space-based MIR nulling interferometer could address. Reliable signal extraction algorithms exist and our results underline the power of the MIR wavelength range for deriving fundamental exoplanet properties from single-epoch observations.}
    

   \maketitle

    \section{Introduction}\label{sec:intro}

Ever since the first detection of an exoplanet orbiting a Solar-like star \citep{Mayor1995AStar}, thousands of exoplanets have been detected from the ground and with dedicated space missions. Ongoing and future space missions, such as the \emph{CHaracterizing ExOPlanet Satellite (CHEOPS)} \citep{cheops2020}, the \emph{James Webb Space Telescope (JWST)}, and the \emph{Atmospheric Remote-sensing Infrared Exoplanet Large-survey (ARIEL)} \citep{tinetti2018}, will focus on characterizing a subset of the known exoplanets and their atmospheres in greater detail. All of these missions rely on investigating exoplanets that transit in front of their host stars, but the vast majority of exoplanets do not transit. Hence, the characterization space is limited and biased towards close-in and large planets. In order to investigate the atmospheric properties of hundreds of terrestrial exoplanets, including dozens that are potentially habitable, optimized large-scale space missions that rely on a direct detection technique will be required. This could be either a single aperture UV/optical/NIR telescope to characterize exoplanets in reflected light \citep[e.g., the \emph{HabEx} and \emph{LUVOIR} concepts;][]{habex2019,luvoir2019}, or a  mid-infrared (MIR) nulling interferometer to probe the thermal emission of exoplanets. In \cite{paper1} we have introduced the \emph{Large Interferometer For Exoplanets (LIFE)} initiative that aims at developing the science objectives, requirements and a concept (incl. a technology development roadmap) for such a space-based MIR nulling interferometer mission.
Of significant importance in the current phase of the \emph{LIFE} initiative is the availability of a simulator environment that, under well-defined assumptions, can assess the performance of a certain interferometer architecture and instrument concept in terms of scientific output. Similarly, a solid understanding of how well a measured signal can be extracted from the simulated data is crucial. 
The longer-term objective is to develop and maintain a flexible framework that incorporates the fundamental properties of various interferometer architectures and includes all relevant noise sources (astrophysical and instrumental). Such a framework can then be used for even  more sophisticated scientific simulations and trade-offs (e.g., prioritization of targets), but also to understand and quantify the impact of technical trade-offs. 

The structure of the paper is as follows: 
in Sect.~\ref{sec:simulation}, we give a brief introduction to the basic principles of nulling interferometry needed to describe the measuring process of \emph{LIFE}. We further introduce the most dominant astrophysical noise terms, explain how the signal-to-noise ratio (SNR) of a measurement can be quantified and provide simulations of an Earth-twin exoplanet seen at a distance of 10 pc as a specific example. In Sect.~\ref{sec:extraction}, we revisit the assumption of the photon-based SNR calculations and present a signal extraction method that enables locating exoplanets in the vicinity of their host stars and estimating their fluxes. In Sect.~\ref{sec:analysis} we investigate how accurately physical properties (such as position and flux and, from this, effective temperature and radius) of simulated exoplanets can be extracted based on single-epoch observations. We do this by re-analyzing parts of the Monte Carlo simulations presented in \cite{paper1}. We discuss the results and conclude in Sect.~\ref{sec:conclusions}.
\fis{Throughout this work, we assume that the measurement is primarily disturbed by the astrophysical noise sources and not by noise emerging from instrumental effect. In Appendix \ref{app:fundamental_noise}, we examine this assumption by presenting estimates for the required instrument stability using an explicit treatment of the instrumental noise.}
    
\section{Signal simulation}\label{sec:simulation}
    
\subsection{Nulling interferometry}
\label{sec:nulling}

     \cite{Bracewell1978DetectingInterferometer} is the first to propose nulling interferometry as a method to detect and characterize exoplanets.
     The general goal of a nulling interferometer is to enable measurements of a faint source - i.e., of an exoplanet - despite the presence of a much brighter source close by.
     The simplest realization, a single Bracewell interferometer, consists of two collector apertures which are separated by a baseline $b$. By combining the collected light coherently, the response of the instrument is a sinusoidal fringe pattern with spacing $\lambda/b$ projected onto the plane of the sky, $\lambda$ being the observing wavelength \citep[e.g.,][]{Bracewell1978DetectingInterferometer,Lay2004SystematicInterferometers}. By adding a $\pi$-phase shift to the beam of one of the collectors, the central minimum of the fringe pattern coincides with the position of the star so that its signal is effectively suppressed, i.e., ``nulled''.
    Nulling the on-axis stellar light is hence a way to optically separate the star light from other nearby sources, such as exoplanets. 
    
    An interferometer has the advantage that it provides higher ``spatial resolution'' than a single-aperture telescope with aperture size $D$: the first positive transmission peak in the fringe pattern is located at $\lambda/2b$, while the spatial resolution of a telescope is $\propto\lambda/D$. Furthermore, in case of a free-flying space interferometer as foreseen for \emph{LIFE}, the baselines are re-configurable and generally much larger (up to hundreds of meters) than the apertures of a single-dish telescope (i.e., $b>>D$).
    
    In a static, monochromatic configuration, the single Bracewell interferometer provides only limited coverage \fis{of the uv-plane} in the form of a single baseline. Increasing this coverage will allow for the retrieval of more complex signals from the measurements. One possible method for enhancing coverage without changing the configuration of the array itself can be achieved by rotation of the array (cf. rotation synthesis, \cite{paresce_science_1997}). By continuously rotating the interferometer around its line of sight, the projected fringe pattern will also rotate, and while the on-axis star remains nulled, any planet in its vicinity moves through the pattern of varying transmission and its signal is modulated.
    The uv-plane coverage can be additionally increased by adding collector apertures to the array and hence raising the number of baselines, or by performing multi-wavelength observations (see Fig.~\ref{fig:analysis_psf}).

    
    Nonetheless, the single Bracewell interferometer is insufficient for the detection of Earth-like exoplanets \citep[e.g.,][]{Angel1997AnPlanets,Defrere2010NullingMissions}\footnote{We refer the reader to \cite{dandumont2020} of recent detection yield estimates for Bracewell interferometers with various aperture sizes.}. Furthermore, because the transmission pattern is symmetric, an ambiguity of $180^\circ$ exists for the position angle of any detected exoplanet.

    Driven by the ultimate goal to detect Earth-like exoplanets, extensive studies of interferometer configurations consisting of more than two apertures led to two major space mission concepts in the early 2000s: the \emph{Darwin} concept led by the European Space Agency (ESA)\citep{Cockell2009Darwin-anPlanets} and NASA's \emph{Terrestrial Planet Finder - Interferometer (TPF-I)} \citep{tpfi2007}. \emph{LIFE} leverages the heritage of these concepts \rs{by rooting its design in analyses performed in the context of the \emph{Darwin} and \emph{TPF-I} missions.}
    
\subsubsection{Beam combination in an X-array configuration} \label{sec:PCA}
\begin{figure}
        \centering
        \includegraphics[width=0.75\linewidth]{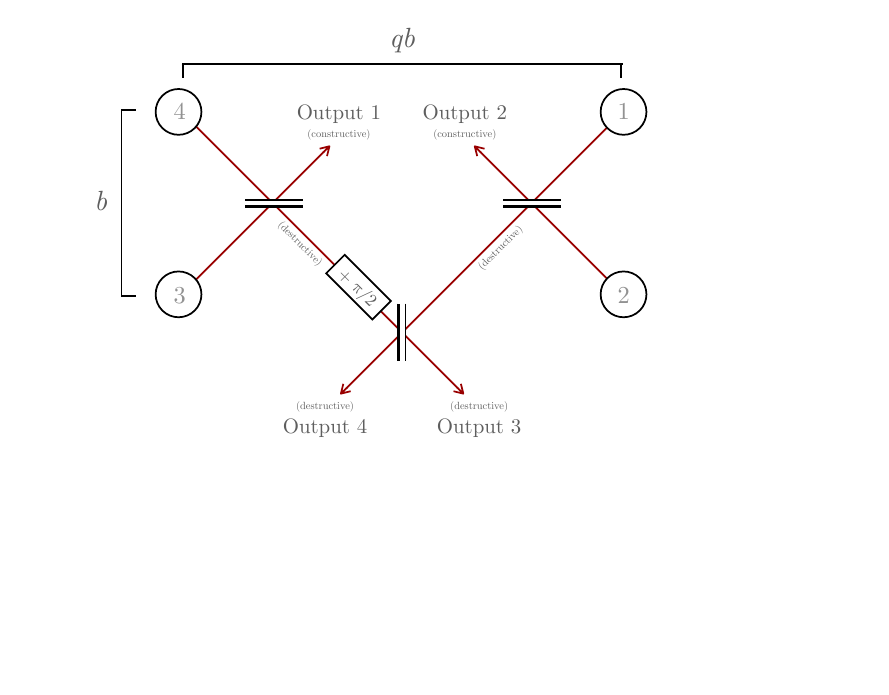}
        \caption{Assumed beam combination scheme of the LIFE array. Four collector spacecraft are located on a 6:1 rectangle with the combiner spacecraft located in the geometric center. The shorter distance between the collectors $b$ is called the nulling baseline, the longer distance is the imaging baseline $qb$. The dual chopped Bracewell beam combination scheme, consisting of an interconnection of two single Bracewell combiners, is shown. We note that the collector spacecraft and the interferometric outputs are numbered for the subsequent calculation and that the spacecraft diameters are not to scale.}
        \label{fig:life_diagram}
    \end{figure}

    For \emph{LIFE} we assume an X-array architecture. It consists of four free-flying collector spacecraft, arranged in a rectangular configuration (see Fig.~\ref{fig:life_diagram}) in a plane perpendicular to the line of sight. Because the spacecraft are free-flying, the baselines can be scaled depending on the desired spatial extent of the fringe pattern. At the moment, the ratio between the length of the long so-called ``imaging baseline'' and that of the short so-called ``nulling baseline'' is chosen to be 6:1  \citep[cf.][]{Defrere2010NullingMissions}. Compared to smaller ratios, the 6:1 configuration is better suited for advanced post-processing techniques aimed at removing a part of the instrumental effects on the measurement called 
    instability noise \citep{Lay2006}. 
    Further investigations and trade-offs are required to validate and confirm this choice for the \emph{LIFE} mission. The fifth spacecraft, in which the beams are combined coherently and the signal is detected, is located in the geometric center of the array. However, this beam combiner spacecraft could either be located in the same plane as the collector spacecraft or, alternatively, it could fly several hundreds of meters above this plane. \rs{This trade-off mostly affects stray light and viewing zone considerations and while it is of no concern for the following analyses, it does significantly affect the field of regard of the mission \citep[cf.][]{Lay2007Planet-findingArchitecture}.}

     The beam combination for such complex array configurations need to be captured by a suitable formalism. The following presents such a formalism as described by \cite{Guyon2013OptimalInterferometers}.

    To calculate the coherent beam combination, a nulling interferometer can be fully described by $N$ apertures, each defined by its respective position and radius $(x_k, y_k, r_k)$ in a two-dimensional plane, where $k\in[1,\dots,N]$. The complex field amplitude generated in the $k$-th aperture by a point source with projected angular offset ($\alpha, \beta)$ from the central line-of-sight is given by
    
    \begin{align}
            V_k = r_k \, e^{\mathrm{i}2 \pi (x_k \alpha +y_k \beta)/ \lambda}\quad,
    \end{align}
    where $\lambda$ is the wavelength. 
    The interferometric combinations can then be represented by a matrix $\vec{\rm U}$, describing the linear mapping from input amplitude vector $\Vec{V}$ to the output amplitude vector
    
    \begin{align}
        \Vec{W} = \vec{\rm U}\Vec{V}.
    \end{align}

    \begin{figure*}[t]
        \centering
        \includegraphics[width=0.98\linewidth]{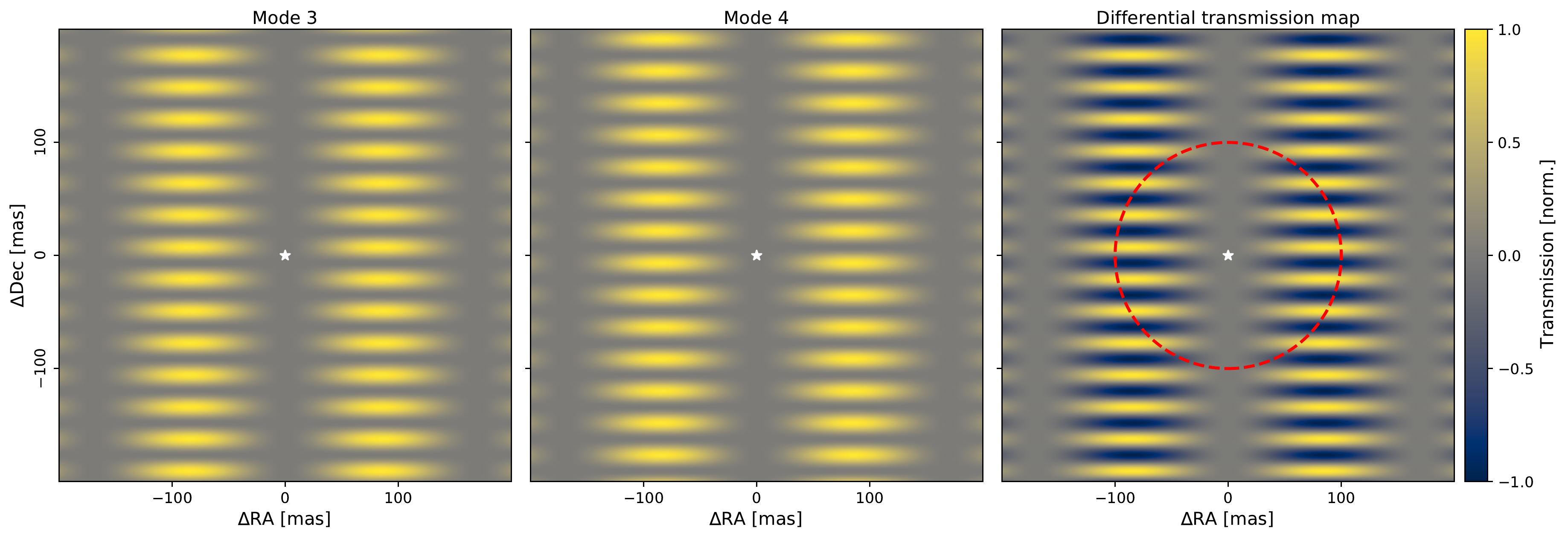}
        \caption{\textit{Left and middle}: Transmission maps $T_\textit{m}$ of \rs{output} 3 and 4 calculated for a wavelength $\lambda = \SI{10}{\micro\meter}$, a nulling baseline $b = \SI{12}{\meter}$ and a 6:1 baseline ratio. These two \rs{outputs} have destructive interference at their vertical center and are used to create the differential map. The axes show the distance from the center in milliarcseconds. The white star indicates the position of the star in the center of the map.
        \textit{Right:} Differential map $T_\mathrm{dif}$, which is the difference between $T_\mathrm{3}$ and $T_\mathrm{4}$. The red dashed line shows the path of a hypothetical planet at \SI{100}{mas} through the differential map as the interferometer array describes a full rotation.}
        \label{fig:transmission_modes}
    \end{figure*} 
    

    Based on the heritage of the \emph{Darwin} and \emph{TPF-I} concepts, for this work we assume that light picked up by the collector spacecraft is merged using a dual chopped Bracewell combiner \citep{Lay2004SystematicInterferometers}. The matrix
    
    \begin{align}
        \vec{\rm U} = \frac{1}{\sqrt{4}}\begin{pmatrix}
                0 & 0 & \sqrt{2} & \sqrt{2}\\
                \sqrt{2} & \sqrt{2} & 0 & 0 \\
                1 & -1 & -e^{i\frac{\pi}{2}} & e^{i\frac{\pi}{2}} \\
                1 & -1 & e^{i\frac{\pi}{2}} & -e^{i\frac{\pi}{2}}
            \end{pmatrix}
    \end{align}
    describes a realizable (matrix is unitary, cf. \cite{Guyon2013OptimalInterferometers}) and lossless (matrix is orthonormal, cf. \cite{Laugier.2020}) implementation of this combiner.
    
    The rows of this matrix reflect the two-stage approach of the beam combination (Fig.~\ref{fig:life_diagram}). In a first stage, the collector apertures are grouped and combined into two single Bracewell combiners. The respective constructive outputs, denoted as Outputs 1 and 2, are captured by the first two rows of the matrix $\vec{\rm U}$, which do not induce any phase delays. The destructive outputs of the first stage are expressed by the $e^{i\pi}$ phase difference between entries one/two or three/four in the last two rows of the combiner matrix.
    In a second stage, these destructive outputs are combined again, with one of the destructive outputs receiving an additional phase shift of $\frac{\pi}{2}$. This produces two complementary nulled outputs that can be used to apply phase chopping to the final output, which reduces the susceptibility of the measurement to instrumental noise effects \citep{Lay2004SystematicInterferometers}. The outputs of this second combination stage are denoted as the destructive Outputs 3 and 4 and are captured by the rows 3 and 4 in the matrix $\vec{\rm U}$. We note that the analyses presented in this paper are solely based on these destructive outputs. 
    
    


\subsubsection{Transmission and differential maps}
    The two-dimensional projected intensity transmission $T_\mathrm{m}$, 
    which describes the amount of signal originating from a point in the sky at position $(\alpha,\beta)$ that transmits to the output  $\textit{m}$, can be described by a combination of the input beams $V_k$ as
    
    \begin{align}
        T_\mathrm{m} = |W_\mathrm{m}|^2 = \bigl\lvert\sum_k U_{\mathrm{m},k} V_k \bigr\rvert^2.
    \end{align}
    
    In the following, the transmission map of one of the interferometric \rs{outputs} is calculated analytically to illustrate the basic principle. In a coordinate system that has its origin in the center of the interferometer array, \rs{the aperture positions $x_k$ and $y_k$} are given as $\pm L$ and $\pm q L$ respectively, with $L$ representing half the nulling baseline $b$, and $q$ being the ratio between the imaging baseline and the nulling baseline. The transmission of \rs{output} 3 as a function of angular separation $(\alpha, \beta)$ is given by
    
    \begin{align}
        T_3 = &\, |W_3|^2 = \bigl\lvert\sum_k U_{3,k} V_k \bigr\rvert ^2 \\
        = &\,\dfrac{1}{4} \,  \bigl\lvert \, e^{\mathrm{i}2\pi (L\alpha + q L \beta)/\lambda}
            - e^{\mathrm{i}2\pi (-L\alpha + q L \beta)/\lambda} \\
            &- e^{\mathrm{i}2\pi (-L\alpha - q L \beta)/\lambda + \mathrm{i} \pi /2}
            + e^{\mathrm{i}2\pi (L\alpha - q L \beta)/\lambda + \mathrm{i}\pi /2}\, \bigr\rvert^2 \\
        =  &\,4 \sin^2 \left(\dfrac{2\pi L \alpha }{\lambda}\right)
            \cos^2 \left(\dfrac{2\pi q L \beta}{\lambda} - \dfrac{\pi}{4}\right) .
    \end{align}
    
    To apply phase chopping, \rs{outputs} 3 and 4 are subtracted from each other. This yields the differential map
    \begin{align}
        T_\mathrm{dif} &= T_3 - T_4 \\
        &= 4 \sin^2 \left(\dfrac{2\pi L \alpha }{\lambda}\right)
            \sin \left(\dfrac{4\pi q L \beta}{\lambda}\right).
    \end{align}

    This map is anti-symmetric with respect to its central point. Thus, any point symmetric emission source (e.g., local zodiacal light, homogeneous exozodi disks) will not transmit through the differential map, but will only contribute to the statistical shot noise \citep[e.g.,][]{Defrere2010NullingMissions}. An example of the two destructive \rs{output} transmission maps $T_3$ and $T_4$ as well as the differential map is shown in Fig.~\ref{fig:transmission_modes}. 
    
    %
    
    
    \begin{figure}[t]
        \centering
        \includegraphics[width=1.\linewidth]{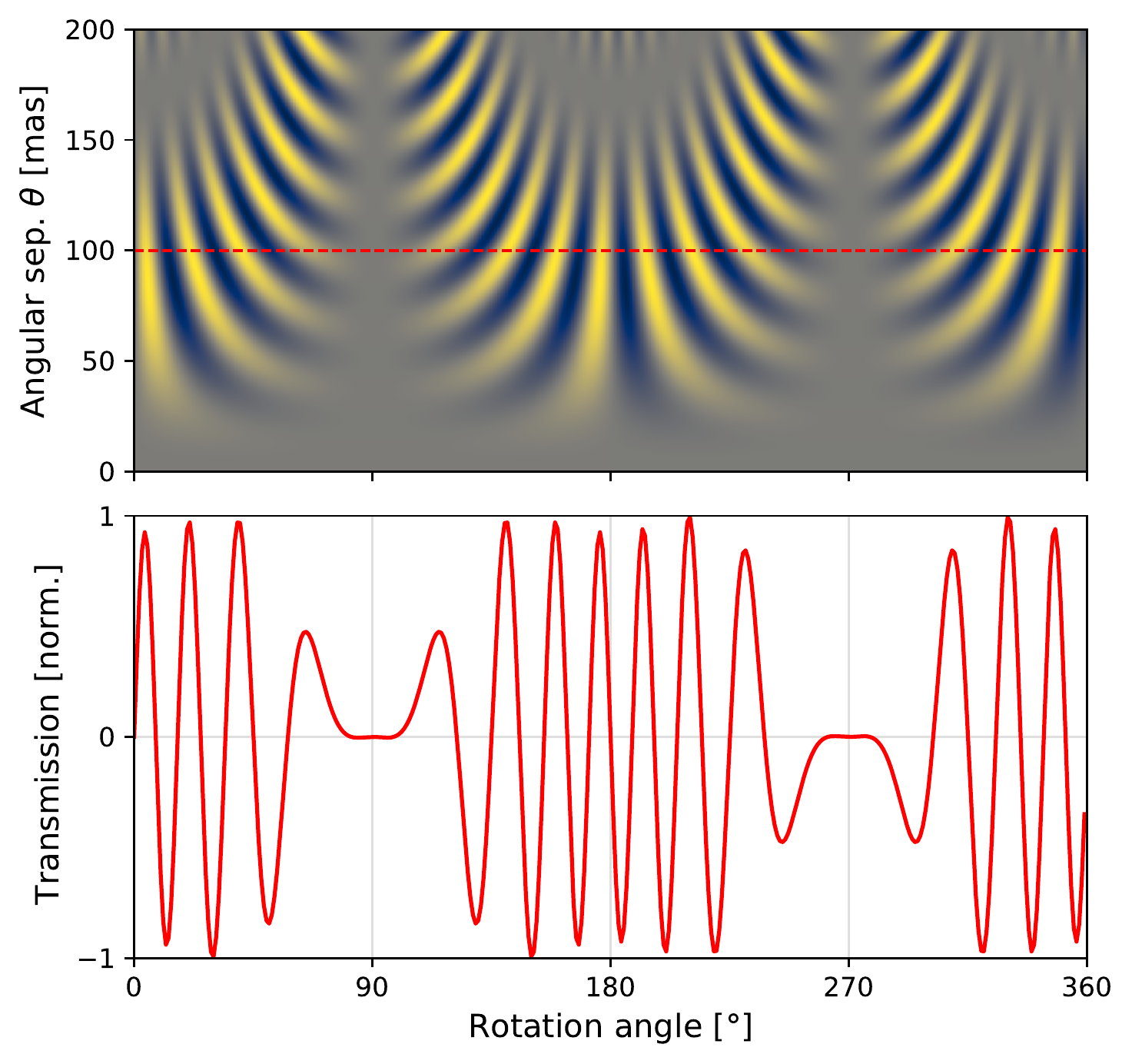}
        \caption{\textit{Top}: Differential map in polar coordinates, normalized from $-1$ (blue) to 1 (yellow). The red dashed line at $\SI{100}{mas}$ angular separation corresponds to the path of the planet shown in the right panel of Fig.~\ref{fig:transmission_modes} for a full rotation of the array. \textit{Bottom}: The normalized signal transmission intensity of the planet as a function of rotation angle.}
        \label{fig:pol_map_and_signal_profile}
    \end{figure}
    
\subsubsection{Signal modulation}\label{sec:signal_transmission}
    To study the signal transmission of a potential planet, the differential map is examined in polar coordinates as shown in the top panel of Fig.~\ref{fig:pol_map_and_signal_profile}. Any (static) source at some angular separation $\theta$ from the central star will move through this map on a horizontal line as the telescope array is rotated by an angle $\phi$. Thus, the normalized signal generated by such a source over one array rotation corresponds to the value of the differential map on a horizontal line through the map at the corresponding radial distance. The normalized signal transmission as a function of rotation angle
    , i.e.,  the modulated signal, of that point source is shown in the bottom panel of Fig.~\ref{fig:pol_map_and_signal_profile}.
    
\begin{figure}
        \centering
        \includegraphics[width=0.95\linewidth]{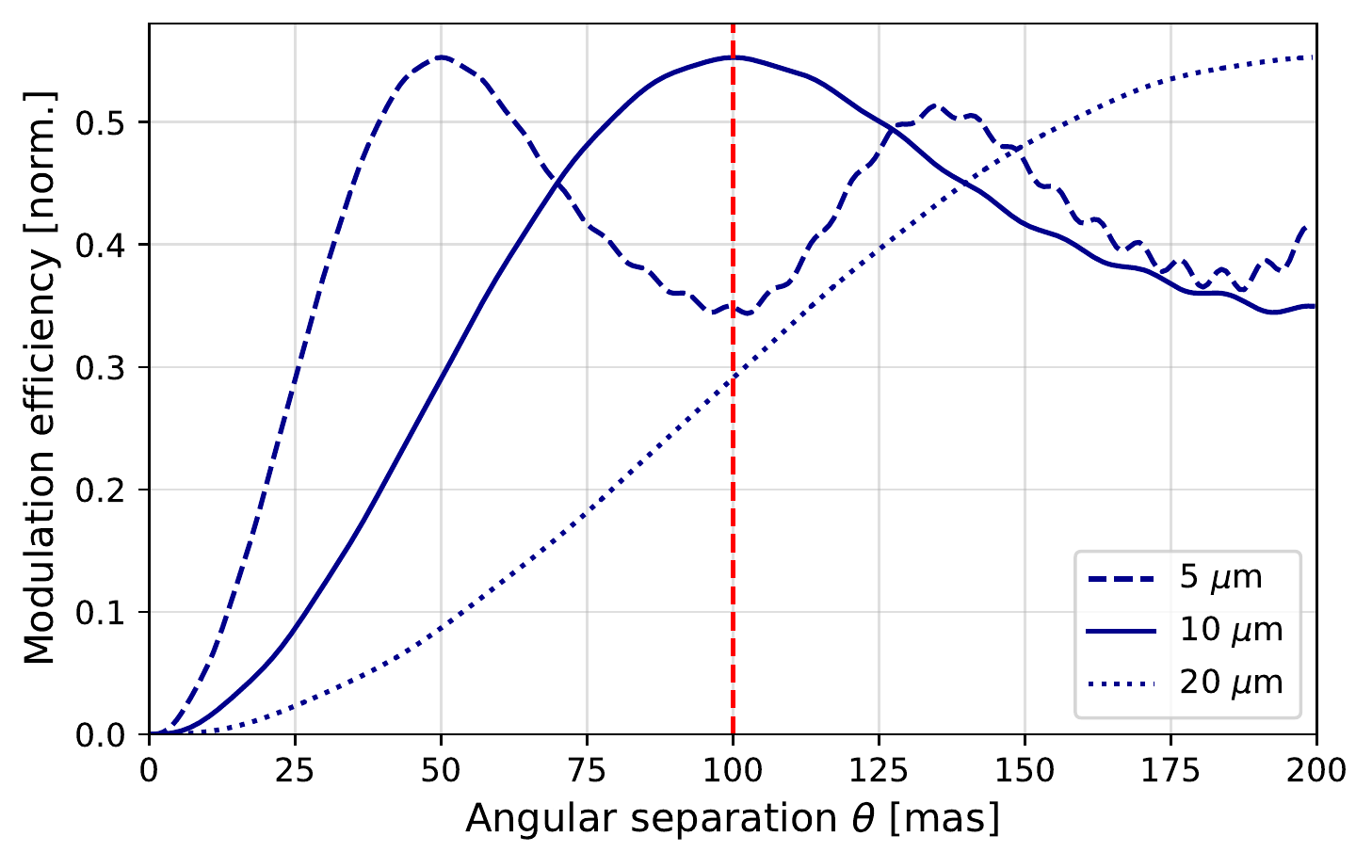}
        \caption{\rs{Modulation} efficiency $\xi(\theta)$ as a function of angular separation for the three wavelengths $\lambda =$ \SI{5}{\micro\meter}, \SI{10}{\micro\meter}, and \SI{20}{\micro\meter}. The vertical red line indicates an angular separation of \SI{100}{mas} corresponding to the separation of the planet shown in the right panel of Fig.~\ref{fig:transmission_modes}.}
        \label{fig:transmission_eff}
    \end{figure}

    The modulation efficiency $\xi$ indicates the part of the incoming signal that can be used for signal extraction and corresponds to the root-mean-square (rms) of the differential map upon a complete array rotation \citep{Lay2004SystematicInterferometers}. The modulation efficiency over a full array rotation as a function of $\theta$ is given by
    
    \begin{align}
        \xi(\theta) = \sqrt{ \langle T_\mathrm{dif}(\theta,\phi)^2 \rangle_\phi}\quad.
        \label{eq:mod_effic}
    \end{align}
    Because the spatial extent of the fringe pattern depends on the baseline configuration of the interferometer and the considered wavelength, this is also true for the modulation efficiency. Figure~\ref{fig:transmission_eff} shows $\xi(\theta)$ for three different MIR wavelengths for the baseline configuration used in Fig.~\ref{fig:transmission_modes}. The curve for $\lambda = \SI{10}{\micro\meter}$, which is based on the differential map shown in Figs.~\ref{fig:transmission_modes} and~\ref{fig:pol_map_and_signal_profile}, has its first peak and global maximum at $\SI{100}{mas}$. 
    For shorter wavelengths the pattern shrinks and the peak transmission is located closer to the star; for longer wavelengths the pattern expands and the peak transmission moves away from the star. The position of the maximum of $\xi(\theta)$ can be calculated numerically as
    
    \begin{align}
        \theta_\mathrm{\xi_{max}} &= 0.59 \, \dfrac{\lambda}{2L}  = 0.59 \, \dfrac{\lambda}{b}\quad,
        \label{eq:max_mod_eff}
    \end{align}   
    which is slightly more than the separation corresponding to the first maximum in the transmission maps at  $\lambda/2b$. 

    \subsection{Astrophysical sources} \label{sec:astro_sources}
        In the following we describe how various relevant astrophysical sources and their photon flux are implemented in \textsc{LIFEsim}.
        
        
    \subsubsection{Host star and exoplanets}
        To first order, thermal radiation emitted by an exoplanet and its host star can be approximated as black body radiation.
        Because we will base our SNR calculation on the principles of photon statistics, we are interested in the received spectral photon flux in units of [\si{ph\,\second^{-1}\meter^{-2}\meter^{-1}}]

        \begin{align}
            F_\lambda(\lambda, T) = \dfrac{2  c}{\lambda^4} \dfrac{1}{e^{hc/\lambda k_\mathrm{B} T}- 1} \dfrac{\pi R^2}{d^2}\quad,
            \label{eq:bb_flux_photons}
        \end{align}
        where $h$ is Planck's constant, $c$ is the speed of light, $k_\mathrm{B}$ is the Boltzmann constant, $T$ and $R$ are the effective temperature and radius of either the planet or the star, and $d$ is the distance from the observer.

        Throughout this paper the star is modeled as a black body with effective temperature $T_\mathrm{s}$ and radius $R_\mathrm{s}$. Limb darkening is not taken into account, and thus the stellar brightness is distributed uniformly across the projected circular disk. Because of the principle of geometric stellar leakage, as described further below in Sect. ~\ref{sec:stellar_leakage}, this is thought to be a conservative approximation.
        
        As one of the objectives of \emph{LIFE} is the atmospheric characterization of (terrestrial) exoplanets by measuring wavelength-dependent absorption features in their emission spectra, modeling their emission with black body radiation is a simplification. Therefore, \textsc{LIFEsim} has the option to select either a simple black body model for the planet or to upload a simulated / observed emission spectrum. In the following subsections we make use of both options. When using a black body, we denote the radius and the effective temperature of the exoplanets with $R_\mathrm{p}$ and $T_\mathrm{p}$. We use a simulated Earth's spectrum for illustrative purposes when calculating the expected SNR of an Earth-twin in Sect.~\ref{sec:methods_earth_twin}. 

    \subsubsection{Stellar leakage}\label{sec:stellar_leakage}
        Even with a perfect nulling interferometer, photons from the observed star ``leak'' through the instrument. If the central null fringe is positioned on the center of the star, the stellar flux is not completely suppressed because of the star's finite size. This phenomenon is described as geometric stellar leakage \citep[e.g.,][]{Lay2004SystematicInterferometers}.
        The photon noise contribution from geometric stellar leakage is determined by applying the transmission maps to the surface area of the star. As such, it depends on the surface brightness and angular size of the star as well as on the baseline configuration.

    \subsubsection{Local zodiacal dust}
        Our Solar System contains diffuse dust that is mostly located near the ecliptic plane. The dust adds a radiation background to any astronomical observation either coming from sunlight that is scattered off the dust particles or as thermal radiation emitted by the dust itself. As \emph{LIFE} works in the MIR, we concentrate solely on the thermal emission. The local zodiacal surface brightness is described by a sky position dependent model developed for the \textsc{DarwinSim} science simulator \citep{Hartog2005TheSimulator}. The parametrization, which is based on data from COBE \citep{Kelsall1998Cloud}, gives the spectral surface brightness
        
        \begin{align}\label{eq:LZdarwinsim}
        \begin{split}
            I_\lambda(\lambda, \lambda_{\mathrm{rel}}, \beta) = &\,\tau \,\left[ B_\lambda(\lambda,T_{\mathrm{eff}}) + A\cdot B_\lambda(\lambda, T_\odot) 
            \left(\dfrac{R_\odot}{\SI{1.5}{AU}}\right)^2\right] \\
            &\cdot 
            \left[
            \dfrac{\pi 
            / \arccos(\cos(\lambda_{\mathrm{rel}}) \cos(\beta))}
            {\sin^2(\beta)
            + 0.36 \cdot \left(\dfrac{\lambda}{\SI{11}{\micro\meter}}\right)^{-0.8} \cos^2(\beta)
            } 
            \right]^{\frac{1}{2}}
        \end{split}
        \end{align}
        where $\lambda$ is the wavelength, $\lambda_{\mathrm{rel}} = \lambda_{\mathrm{ecl}} - \lambda_{\mathrm{ecl, \sun}}$ is the ecliptic longitude relative to the Sun, $\beta$ is the ecliptic latitude, $\tau = \SI{4e-8}{}$ is the optical depth towards the ecliptic poles, $B_\lambda(T)$ is the Plank function, $T_{\mathrm{eff}} = \SI{265}{\kelvin}$ is the effective temperature of the local zodiacal dust cloud at \SI{1}{AU}, $T_\odot = \SI{5778}{\kelvin}$ is the effective temperature of the Sun, $A = 0.22$ is the near-IR dust albedo, and $R_\odot$ is the radius of the Sun.
        \rs{Since the local-zodiacal light received by the interferometer is diffuse, it cannot be brought to interfere destructively. Therefore, it is not possible to remove the local-zodi emission by virtue of the rotating nulling interferometer. Subsequently, the most effective way of minimizing the noise at long wavelengths is to observe in directions where the local-zodi emission is weak \fis{(see Appendix \ref{app:localzodi})}.}
        
       
    
        Compared to the original COBE data, the \textsc{DarwinSim} model overestimates the flux from the zodiacal dust in the $\SI{6}-\SI{20}{\micro\meter}$ range and for a line of sight with a relative latitude of more than $90^\circ$ from the Sun by $10-20\,\%$. For wavelengths below \SI{6}{\micro\meter} the difference increases to a factor of 2-3 at \SI{3}{\micro\meter}. However, at these  wavelengths the photon noise contribution by the  zodiacal dust is several orders of magnitude lower than the photon noise from the stellar leakage as we will show below in Sect.~\ref{sec:SNR_calc}.
        
      
\subsubsection{Exozodiacal dust}
        Exozodiacal dust clouds (``exozodis'' for short) are the equivalent to the local zodiacal dust in other stellar systems. Even if the dust density is as low as in the Solar system, integrated over the full field-of-view the MIR radiation emitted by the dust is two to three orders of magnitude higher than that of an Earth-like planet \citep[e.g.,][]{Defrere2010NullingMissions} and can thus significantly affect the integration time required to detect (terrestrial) exoplanets. 
        
        The exozodiacal dust distribution and its surface brightness are simulated according to the model described by \cite{Kennedy2015EXO-zodiInterferometer}. It assumes a dust distribution in exoplanetary systems similar to the one in the Solar system, except for a global scaling factor $z$, the number of \textit{zodis}\footnote{A probability distribution for $z$ was derived by the HOSTS survey carried out at the Large Binocular Telescope (LBTI) \citep{Ertel2020TheSurvey}.}. The model assumes that the dust emission is optically thin and the dimensionless, face-on surface density $\Sigma(r)$ is approximated by a power law distribution.
        The emission is modeled as black body radiation with the temperature-radius profile of the disk following the equilibrium temperature
        
        \begin{align} \label{eq:ez_temp_dist}
            T(r) = \SI{278.3}{\kelvin} \cdot L_\mathrm{s}^{\frac{1}{4}}\, r^{-\frac{1}{2}}
        \end{align}
        where $r$ is given in AU and $L_\mathrm{s}$ in Solar luminosities.
        The spectral surface brightness of the disk as seen face-on is then given by

        \begin{align} \label{eq:ez_surface_brightness}
            I_\lambda(\lambda, r) = \Sigma(r) \, B_\lambda(\lambda, T(r)).
        \end{align}
        Due to the high temperature and surface density close to the star, much of the emitted radiation originates from the central regions. 
        It is important to note that the given definition of the zodi level $z$ is coupled to the surface density and not to the spectral flux or the total luminosity of the disk. For constant $z$, the total emitted radiation at a given wavelength, as well as the luminosity of the disk, scale with the surface area of the disk and thus with the stellar luminosity $L_{\mathrm{disk}} \propto r_0^2 \propto L_\mathrm{s}$. For high-luminosity stars, which are expected to have larger exozodi disks in this model, the total emitted flux by the disk is thus also larger than for low-luminosity stars.
        
        At the moment, the exozodi dust disks are assumed to be homogeneous and symmetric around the star and always viewed face-on. Therefore, they only contribute to the shot noise (see, Sect.~\ref{sec:SNR_calc} below). The influence of asymmetric or clumpy exozodiacal disks viewed at an inclination $i>0^{\circ}$ was investigated in \cite{Defrere2010NullingMissions} and \cite{Defrere2012} and we discussed these points and their potential impact on exoplanet detection yield in \cite{paper1}.

\subsection{Signal-to-noise ratio calculations}\label{sec:SNR_calc}
\subsubsection{Astrophysical noise sources}\label{sec:SNR_calc_astro}
    The astrophysical sources located within the field-of-view of the collector apertures create a scene that is described by the surface brightness $I(\Vec{\theta}, \lambda)$. It depends on the position $\Vec{\theta}$ within the field-of-view and on the wavelength, and is here assumed to be time-independent.
    
    When the photons emitted by the different sources reach the interferometer, they are detected in one of the interferometric \rs{outputs}. 
    The collected signal $S_\mathrm{m}$ by \rs{output} $\textit{m}$, described by transmission map $T_\mathrm{m}$, after some integration time is given by the integral of the total surface brightness $I(\Vec{\theta}, \lambda)$ superimposed with the transmission map over the complete field-of-view
    
    \begin{align} \label{eq:signal_measurement}
        S_\mathrm{m}(\lambda) = \int T_\mathrm{m}(\Vec{\theta}, \lambda) \,I(\Vec{\theta}, \lambda)\, t\, A\, \eta \,\mathrm{d\Omega}
    \end{align}
    with $t$ the integration time, $A$ the total collecting area, and $\eta$ a detection efficiency factor combining multiple effects. The efficiency factor $\eta$ and the size of the field-of-view are generally also wavelength-dependent. In this work, $\eta$ is the product of the quantum efficiency of the detector $\eta_\mathrm{QE}$ and the overall instrument throughput $\eta_\mathrm{t}$.
    Until a more concrete instrument design is available, it is assumed that both parameters are wavelength-independent. As motivated in more detail in \cite{paper1}, we chose $\eta_\mathrm{QE}=0.7$ and $\eta_\mathrm{t}=0.05$ as default values. The effective field-of-view is assumed to be $\mathrm{FoV} = \lambda/D$ in diameter as we assume the light will be coupled into single-mode fibers.

    The spectral signal generated by a single planet, approximated as a point source at location $\Vec{\theta}_\mathrm{p}$, is thus given by
    
    \begin{align} \label{eq:signal_planet}
        S_\mathrm{p,m}(\lambda) = T_\mathrm{m}(\Vec{\theta}_\mathrm{p}, \lambda) \,F_\mathrm{p}(\lambda)\, t\, A\, \eta
    \end{align}
    where $F_\mathrm{p}(\lambda)$ is the planet flux.
    Over a full array rotation, the quadratic mean of the modulated signal $S_\mathrm{p} = S_\mathrm{3,p} - S_\mathrm{4,p}$ detected from the planet is given by
    
    \begin{align}
        \sqrt{\langle S_\mathrm{p}^2(\lambda)\rangle} = \xi(\lambda, \theta_\mathrm{p}) \, F_\mathrm{p}(\lambda)\, t\, A\, \eta \label{eq:quad_mean_signal}
    \end{align}   
    where $\xi(\lambda,\theta)$ is the wavelength dependent modulation efficiency at angular separation $\theta$ from the star (see, Eq.~\eqref{eq:mod_effic}).
    
    If only rotationally symmetric sources, such as the central star, the local zodiacal background or a homogeneous, face-on exozodiacal dust cloud are considered, the detected signal does not depend on the rotation angle of the array. If the source is only point-symmetric, such as a smooth exozodiacal disk with some inclination $i >0^{\circ}$, the signal per \rs{output} varies with rotation angle, but is always equal for the two destructive \rs{outputs} 3 and 4. Thus, point-symmetric sources do not contribute to the modulated signal. 
    
    However, as \cite{Mugnier2006DataMission} pointed out, the linear combination of detected signals is an incoherent combination, since it is performed numerically after detection (i.e., in post processing). Removing the contribution of symmetrically distributed sources by incoherent combination is therefore only effective for the signal part of the data, but the sources still contribute to the statistical noise.
    Hence, the SNR per spectral wavelength bin, denoted as SNR$_\lambda$ and defined as the ratio between detected exoplanet photons and detected photons from the various noise terms, is calculated as
    
    \begin{align}
        \mathrm{SNR}_\lambda =
        \dfrac{
        \int
        \sqrt{\langle S_\mathrm{p}^2(\lambda)\rangle} \,\mathrm{d}\lambda
        }
        {
        \sqrt{
        2\int
        \, (S_\mathrm{sym, 3}(\lambda)
        + \, \sqrt{\langle S_\mathrm{p, 3}^2(\lambda)\rangle}) \,\mathrm{d}\lambda
        }
        }\quad,
    \end{align}
    where $S_\mathrm{sym, 3}$ is the contribution from symmetric background sources to the signal in \rs{output} 3 and $S_\mathrm{p, 3}$ is the signal of the planet only\footnote{For the computation it does not matter whether we use \rs{output} 3 or \rs{output} 4 in the denominator. The factor of 2 in front of the integral ensures that the contributions from both \rs{outputs} are considered.}. The integral runs over the wavelength range covered by the wavelength bin.
    
    Following \cite{Lay2004SystematicInterferometers}, for a measurement across several wavelength bins and under the assumption that the noise is uncorrelated between the bins, the integrated SNR over the full wavelength range, or summed over all wavelength bins, is given by 
    
    \begin{align}\label{eq:methods_snr_integration}
        \mathrm{SNR_{tot}} &=  \sqrt{\sum_\lambda \mathrm{SNR_\lambda^2}}
        \propto \sqrt{t\,A\,\eta}\quad ,
    \end{align}
    which scales, as expected, with the square root of integration time, area and detection efficiency. 

    In \textsc{LIFEsim} each noise term from the different sources is calculated individually to avoid numerical discretization errors, as the angular extents of the sources have different scales. For the local and exozodiacal dust, the transmission map as well as the surface brightness distribution are calculated and integrated over a two-dimensional artificial image covering the full field-of-view defined by the instrumental parameters. For the stellar leakage, it is only integrated over the solid angle covered by the stellar disk, but with much higher resolution. The planet signal transmission efficiency is calculated analytically using Eq.~\eqref{eq:signal_planet}. All terms are calculated for each pre-defined wavelength bin, with varying angular extent of the field-of-view and possibly varying bin width, and are then combined to an integrated signal-to-noise ratio as given in Eq.~\eqref{eq:methods_snr_integration}.
    
    The calculated SNR depends strongly on the baseline of the array. The mean modulated planet signal (Eq.~\eqref{eq:quad_mean_signal}) scales with the modulation efficiency, which in turn depends on the baseline as demonstrated in Sect.~\ref{sec:signal_transmission}. For the contribution from background sources, the amount of stellar leakage contribution depends on the broadness of the central null. This broadness of the null in the transmission map is also governed by the baseline configuration. Thus, maximizing the SNR by changing the length of the baselines poses a trade-off between the amount of planetary signal received over the amount of stellar leakage affecting the measurement. 
    Ideally, to reduce the integration time required for detection, the baselines of the interferometer should thus be chosen such that the SNR across the full wavelength range is high for projected separations that are of major interest. For instance, maximizing the chances of detecting an Earth-like planet orbiting within the habitable zone (HZ) around nearby stars can be achieved by maximizing the SNR across the projected habitable zone. 
    As discussed in \cite{paper1}, for  \emph{LIFE} it is currently assumed that the baselines can indeed be reconfigured depending on the target star where the minimum and maximum separation between two spacecraft is \SI{10}{\meter} and \SI{600}{\meter} respectively. 

\subsubsection{Instrumental noise terms} \label{sec:instrumental_noise}
    \rs{Instrument perturbations such as intensity variations and optical path difference (OPD) errors can degrade the null and lead to additional stellar leakage as well as instability noise \citep{Lay2004SystematicInterferometers, Defrere2009DetectionPar}.} 
    \rs{Similarly,} detector noise (such as, e.g., dark current) and thermal background noise from the instrument \rs{can also deteriorate the measurement}. 
    \rs{In the current version of \textsc{LIFEsim}}, it is assumed that the instrument will be designed such that photon shot noise \rs{will dominate over these additional instrumental noise sources. Appendix \ref{app:fundamental_noise} presents a preliminary analysis of the aforementioned instrumental effects and constrains the maximum level of allowed perturbations to operate in this fundamental noise limited regime.} 

\subsection{A simple case study: Earth-twin at 10 pc}\label{sec:methods_earth_twin}
    
    We apply the above described SNR calculation to the example of an Earth-like planet located around a Sun-like star at \SI{10}{pc}, with the aim to illustrate the effects of the different noise sources on the SNR and their wavelength dependence. 
    Instrument parameters and properties of the simulated sources are given in Table\,\ref{tab:parameters}. 
    
\begin{figure}[ht]
        \centering
        \includegraphics[width=1.\linewidth]{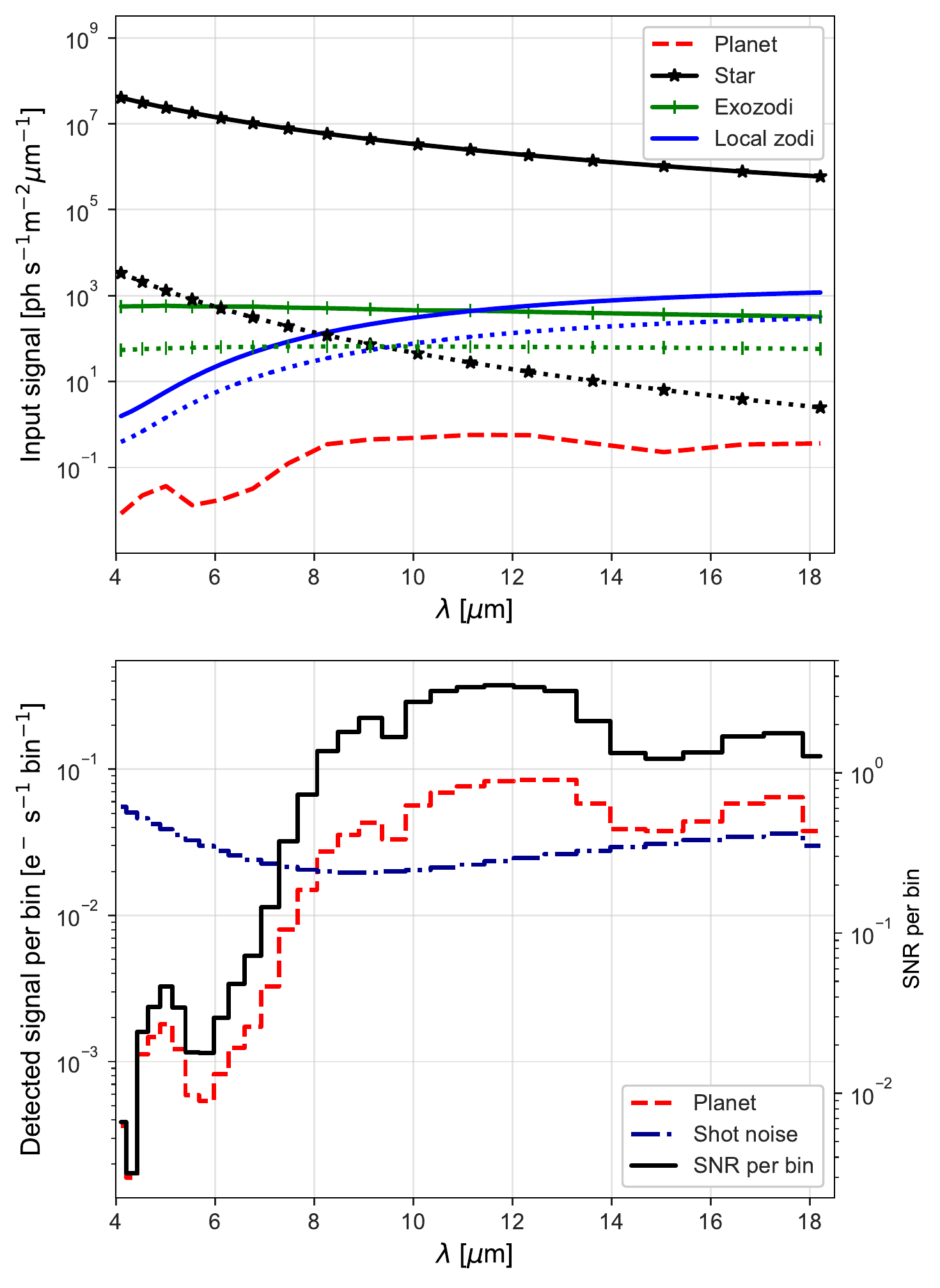}
        \caption{\textit{Top}: Comparison of the various input signals in our example of an Earth-like planet at \SI{10}{pc}. The solid lines indicate the total flux received from the respective sources: planet (red), star (black), exozodiacal dust (grey) and local zodiacal dust (blue). The dotted lines with the same color code indicate the fraction of the flux that contributes to the shot noise. \textit{Bottom}: Detected planet signal (red line) and total shot noise (blue dashed line) averaged over each wavelength bin and for an integration time of \SI{200000}{\second} (values refer to the y-axis on the left). The resulting spectral SNR is shown as the black line and is calculated as described in Sect.~\ref{sec:SNR_calc} (values refer to the y-axis on the right). Integrated over the full wavelength range one obtains an SNR$\approx$9.7.}
        \label{fig:methods_earth_twin_fluxes_and_snr}
    \end{figure}

    Figure~\ref{fig:methods_earth_twin_fluxes_and_snr} (top panel) shows the total fluxes received from the different sources as solid lines. The ``leakage'' terms, i.e., the part of the flux that is not nulled, are shown as dotted lines. These leakage terms contribute to the shot noise. It can be seen that at short wavelengths the stellar leakage dominates strongly, while for longer wavelengths the emission from the local zodiacal dust contributes most to the shot noise. 
    
    In the bottom panel of Fig.~\ref{fig:methods_earth_twin_fluxes_and_snr} we show the detected signal from the planet (averaged over each spectral bin) as well as the shot noise and the SNR for the parameters listed in Table~\ref{tab:parameters}. 
    Integrating over the full wavelength range gives an SNR$\approx$9.7. The maximum SNR per bin of $\approx$3 is reached around \SI{11}{\micro\meter}. For wavelengths shorter than \SI{8}{\micro\meter} the SNR decreases rapidly, due to a decrease in planet signal and the increasing shot noise due to stellar leakage. 
    
    \begin{table}[ht]
    \centering
    \caption{Instrumental parameters and properties of the simulated sources for the SNR calculation of an Earth-like planet at 10 pc.}
        \begin{tabular}{@{}lll@{}}
        \hline\hline 
        Parameter                         & Value                      & Description                                                  \\ \hline
        $D$                               & $\SI{2}{\meter}$           & Aperture diameter                                           \\
        $t$                               & $\SI{200000}{\second}$      & Integration time                                             \\
        $\eta_{QE}$                       & 0.7                       & Quantum efficiency                                           \\
        $\eta_t $                         & 0.05                       & Instrument throughput                                           \\
        FoV                               & $\lambda / D$              & Field-of-view ($\sim1''@ \lambda = \SI{10}{\micro\meter}$) \\
                                          & $4 - 18.5 \,\si{\micro\meter}$ & Wavelength range                                              \\
        $R$                               & 20                         & Spectral resolution ($\lambda/\Delta\lambda$)\tablefootmark{a}                                          \\
        $b$                               & $\SI{15}{\meter}$          & Nulling baseline\tablefootmark{b}                                               \\
        $r$                               & $6:1$                      & Array baseline ratio                                              \\
        $d$                               & $\SI{10}{pc}$              & Target distance                                              \\
        $\theta_\mathrm{p}$                               & $\SI{100}{mas}$              & Planet-star separation                                              \\
        $R_\mathrm{p}$                    & $1\,R_\oplus$         & Planet radius                                                \\
        $T_\mathrm{p, 0}$                    & $\SI{285}{\kelvin}$        & Planet surface temperature\tablefootmark{c}                                           \\
        $R_\mathrm{s}$                    & $1\,R_\sun$          & Stellar radius                                               \\
        $T_\mathrm{s}$                    & $\SI{5778}{\kelvin}$       & Stellar effective temperature                                          \\
        $\lambda_{\mathrm{rel}},\, \beta$ & $135^\circ,\, 45^\circ$             & Ecliptic coordinates                                       \\
        $z$                               & $3$ & Level of exozodi emission\tablefootmark{d}                                     \\ \hline
        \end{tabular}
    \label{tab:parameters}
    \tablefoot{\fis{The table corresponds to Figure \ref{fig:methods_earth_twin_fluxes_and_snr}.} \tablefoottext{a}{Spectral resolution is assumed to be constant across the full wavelength range, such that the bin width $\Delta\lambda$ increases for larger wavelengths. For the parameters listed here, this results in 31 spectral bins.}\tablefoottext{b}{The baseline is set according to Eq.~\eqref{eq:max_mod_eff} \rs{evaluated at \SI{15}{\micro\meter}}.}\tablefoottext{c}{In Fig.~\ref{fig:methods_earth_twin_fluxes_and_snr}, instead of assuming blackbody emission for the planet, we used the radiative transfer atmospheric model code \texttt{petitRADTRANS} to compute an MIR emission spectrum corresponding to an average cloud-free modern Earth spectrum \citep[][]{konrad2021large}}\tablefoottext{d}{This value corresponds to the median of the best-fit nominal model derived from the HOSTS survey \citep{Ertel2020TheSurvey}.}}  
    \end{table}

    \section{Signal extraction}\label{sec:extraction}
    The approach and the simulated spectrum presented in the previous section are based on photon counting statistics. It was implicitly assumed that the signal extraction can be performed with sufficient accuracy. 
    
    In practice, data processing and signal extraction algorithms will be of great importance, in particular if a mission like \emph{LIFE} features an initial search phase\footnote{The baseline mission scenario features a 2.5 year search phase for detecting previously unknown exoplanets, followed by a 2.5-3.5 year characterization phase \citep{paper1}.}: in addition to separating the planetary signal from the various noise sources, it will be crucial to accurately derive physical characteristics of the exoplanet (e.g., radius, effective temperature and separation from the host star) from single epoch data in order to identify and rank-order the most interesting objects for in-depth follow-up observations during the characterization phase. For both the \textit{Darwin} and the \textit{TPF-I} missions, multiple signal analysis algorithms were proposed. \cite{Draper2006Interferometer} presented an algorithm which was based on an advanced correlation process. For \textit{Darwin}, \cite{Mugnier2006DataMission} and \cite{Thiebaut2005MaximumInterferometer} presented a signal extraction scheme based on the maximum-likelihood method (MLM). While in the meantime significant progress has been made in many areas of data post-processing, especially thanks to machine learning based approaches, the before-mentioned methods still offer a straight-forward framework to deal with the unique characteristics of nulling interferometry with respect to the modulation of the planet signal.
    
    In the following subsections the maximum-likelihood method will be described, modified, and applied to simulated \emph{LIFE} data. The goal is to find a metric to quantify the robustness of an exoplanet detection and compare the extracted SNR to that from the photon statistics used in the previous section. Additionally, the signal extraction for multi-planet systems will be investigated. Furthermore, we will revisit the Monte Carlo simulations presented in \cite{paper1} and quantify how accurately we can expect exoplanet properties to be determined from single epoch data during the search phase. 
\subsection{Maximum-likelihood method}\label{sec:MLM}
    As described in Sect.~\ref{sec:SNR_calc}, the signal measurement with a nulling interferometer can be described as a noisy time series. Following \cite{Mugnier2006DataMission} and \cite{Thiebaut2005MaximumInterferometer}, it is modeled as
    \begin{align}
        A_{t,\lambda} &= F_{\mathrm{p},\lambda} \, T_{t, \lambda}(\Vec{\theta}_\mathrm{p}) + n_{t, \lambda}
    \end{align}
    where $A_{t,\lambda}$ is the recorded amplitude at time $t$ and for effective wavelength $\lambda$, $T_{t, \lambda}(\Vec{\theta}_\mathrm{p})$ is the response of the instrument (i.e., the value of the differential map $T_\mathrm{dif}$) at the location of the planet $\Vec{{\theta}}_\mathrm{p}$ as a function of time and wavelength, $F_{\mathrm{p},\lambda}$ is the discretized planet flux and $n_{t, \lambda}$ denotes the noise, which is assumed to be spectrally independent normal noise and whose variance $\sigma^2(t,\lambda)$ can be estimated from the data. In a first step and as a simplification, only the case with one (detectable) planet per system is considered to derive the signal analysis; afterwards, the approach will be extended to multiple planets.
    
    \begin{figure}[t]
        \centering
        \includegraphics[width=1.\linewidth]{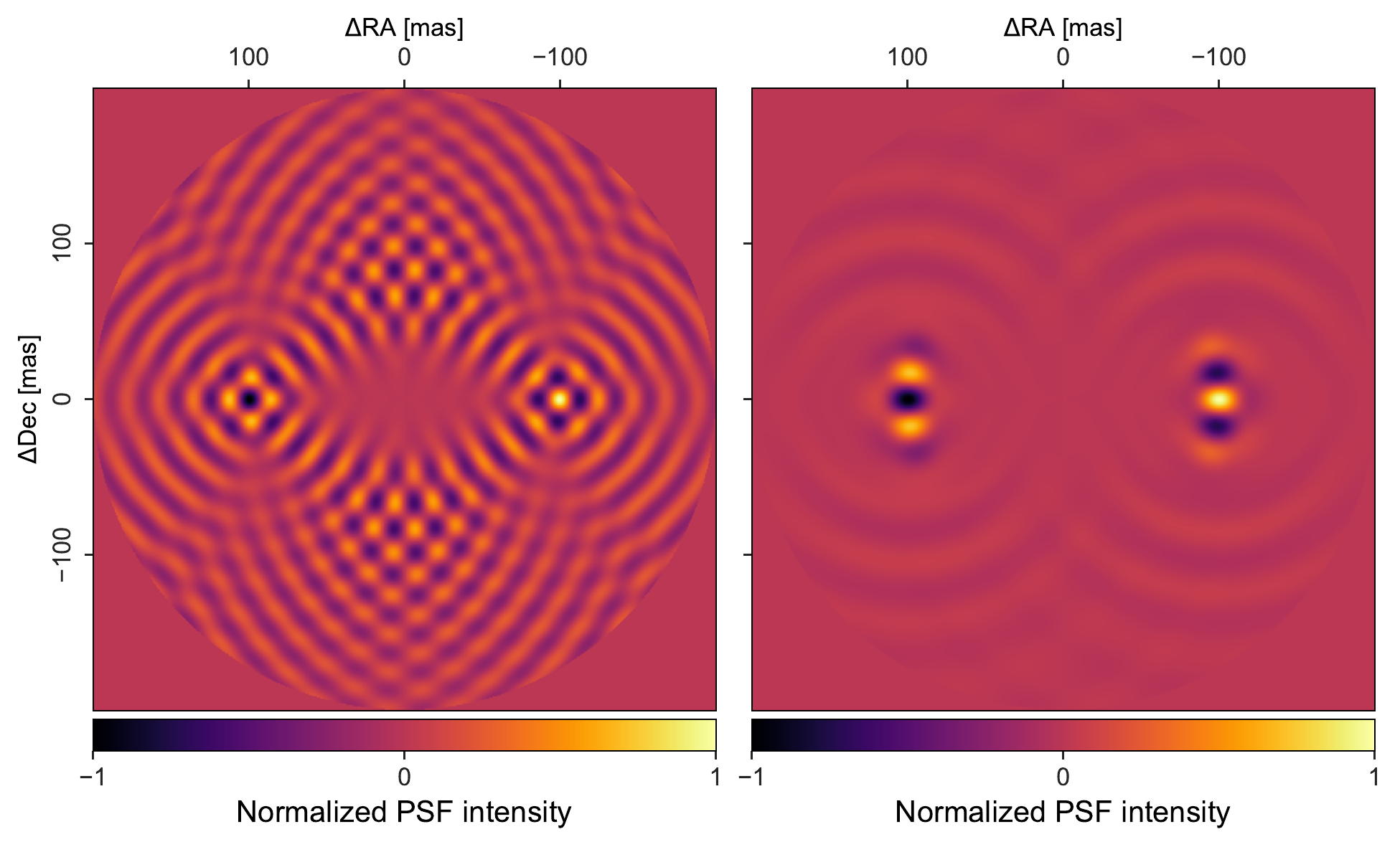}
        \caption{\textit{Left:} Cross-correlation of the modulated signal of a noiseless point source located at [-100, 0] mas with the transmission map at a single wavelength. The resulting image can be understood as the demodulated signal function of such a point source at the considered wavelength. The image is anti-symmetric with respect to the image center. \textit{Right:} The combined cross-correlation across the full wavelength range considered for the ideal point source. Considering a broader wavelength range reduces the side lobes and allows for a better location of the point source.}
        \label{fig:analysis_psf}
    \end{figure}
    The maximum likelihood method enables estimating the planet position $\Vec{\theta}_\mathrm{p}$ and spectral flux $F_\mathrm{p}(\lambda)$ from the modulated signal 
    by searching $(\hat{\Vec{\theta}}_\mathrm{p}, \hat{F}_\mathrm{p})$ that maximize the likelihood $L(\Vec{\theta}_\mathrm{p}, F_\mathrm{p})$.
    Maximizing the likelihood is equivalent to minimizing the negative log-likelihood or cost function
    \begin{align}\label{eq:cost_function}
        &J(\Vec{\theta}_\mathrm{p}, \Vec{F}_\mathrm{p}) =  \sum_{t,\lambda}
        \dfrac{|A(t,\lambda) - F_\mathrm{p}(\lambda)\, T(t, \lambda, \Vec{\theta}_\mathrm{p})|^2}
        {\sigma^2(t,\lambda)}\quad ,
    \end{align}
    which measures the discrepancy between the data $A$ and the model of the data for the estimated set of parameters $(\Vec{\theta}_\mathrm{p}, \Vec{F}_\mathrm{p})$, based on the differential map $T$. $\Vec{F}_\mathrm{p}$ denotes the set of flux values at all measured wavelengths.
    %
    It has been shown in the previously mentioned papers that for a given planet position $\Vec{\theta}_\mathrm{p}$, the optimal spectral flux is obtained by maximizing $J$ (as defined in Eq.~\eqref{eq:cost_func_vec}) with respect to all $F(\lambda$):
    \begin{align}\label{eq:cost_opt_criterion}
        \dfrac{\partial J(\Vec{\theta}_\mathrm{p}, \Vec{F}_\mathrm{p})}{\partial F(\lambda)} = 0 \quad \forall \lambda .
    \end{align}
    To ensure the result is physical, a positivity constraint has to be introduced on the estimated flux.
    The most likely planet position is then found by globally maximizing the cost function over a grid by inserting the optimized flux values back into the cost function.
     The estimation of the planet position and wavelength dependent flux can be further improved if an agreement of the estimated parameters with a priori information is enforced \citep{Thiebaut2005MaximumInterferometer}. This can be done by introducing a regularization term $J_\mathrm{prior}$, which estimates the roughness of a sampled planet spectral flux:
    \begin{align}\label{eq:Jprior}
        J_\mathrm{prior}(\Vec{\theta}_\mathrm{p}, \Vec{F}_\mathrm{p}) 
        = \mu \sum_{\lambda} 
        \left( \dfrac{\partial^m F_\mathrm{p}(\lambda)}{\partial \lambda^m} \right)^2 
        \quad,
    \end{align}
    where $\mu$ is a parameter allowing to tune the relative weight of regularization. The $m$-th order derivative is computed by finite differences \citep{Thiebaut2005MaximumInterferometer}. To smooth the estimated flux, only second order derivatives are taken into account ($m$=2). The calculation of the optimal flux values and the cost function with and without regularization is further outlined in Appendix\,A. In accordance with the previous works, we denote by $J'$ the cost function that is optimized with respect to the fluxes, and by $J''$ the cost function that has an additional positivity constraint on the flux.
    The cost functions can be computed on a grid of possible initial planet positions spanning the full or only a part of the field-of-view. The most likely planet position $\Vec{\hat{\theta}}_\mathrm{p}$ is then the location of the maximum of $J''(\Vec{\theta}_\mathrm{p})$ on the grid.
    
    It is noted here that the cost function is closely related to the cross correlation of an ideal modulated signal generated by a noiseless point source with the template functions (the transmission maps). The cross correlation across a two-dimensional grid gives the demodulated signal function of the interferometer \citep{Defrere2010NullingMissions}. This is shown in Fig.~\ref{fig:analysis_psf} for a single wavelength and for the combination of multiple wavelength bins by addition of the maps.

\subsection{Detection criterion}\label{sec:analysis_detection_criterion}
    To create a link between the outcome of the maximum likelihood signal extraction method and a quantitative detection criterion we further investigate the noise propagation and the probability of false alarm in the following. 
    As introduced in Sect.~\ref{sec:SNR_calc}, and following the definition of \cite{Mugnier2009OptimalImaging}, the SNR of the estimated planet signal at a single wavelength is given by
    \begin{align}
        \mathrm{SNR_\lambda}(\Vec{\theta}_\mathrm{p}) = \mathrm{SNR}[\hat{F}_\mathrm{p}(\Vec{\theta}, \lambda)] &\hat{=} 
        \dfrac{\hat{F}_\mathrm{p}(\Vec{\theta}, \lambda)}{\sigma(\hat{F}_\mathrm{p}(\Vec{\theta}, \lambda))}.
    \end{align}
    with $\sigma(\Vec{\hat{F}}_\mathrm{p}(\Vec{\theta}))$ the standard deviation of the estimated flux.
    Given the high photon flux values of the astrophysical sources introduced in Sect.~\ref{sec:astro_sources} (cf. Fig.~\ref{fig:methods_earth_twin_fluxes_and_snr}), the noise in a measurement can be approximated by a normal distribution.
    To get an estimated value for the integrated SNR, the individual wavelength bins could be combined naively by
    \begin{align} \label{eq:analysis_snr_int_simple}
       \mathrm{SNR_{tot}} = \sqrt{ \sum \mathrm{SNR_\lambda}^2}
                            = \sqrt{ \sum \left( \hat{F}_\mathrm{\lambda}/\hat{\sigma}_\lambda \right)^2},
    \end{align}
    as is done to calculate the integrated SNR from the predicted photon noise in Sect.~\ref{sec:SNR_calc}. However, this does not take into account the random nature of the measurement, i.e., in reality, even repeating the exact same observations will result in a slightly different estimated SNR. 

    \cite{Flasseur2020PACOSpectrographs} presents a method that allows us to derive a detection criterion combining measurements at multiple wavelengths with uncorrelated noise on the basis of the cost function for direct imaging methods. With some modifications, this approach can also be applied to nulling interferometry.
    We note that the dominant and normally distributed background noise in the modulated signal propagates as a linear combination to the SNR$_\lambda$ maps. This means that in the absence of point sources the distribution of SNR$_\lambda$ values over a large enough field-of-view also follows a normal distribution. Thus, the random values of the combined cost function $J'$ map are expected to follow a $\chi^2$-distribution $p(J') = \chi^2_L$ with $L$ degrees of freedom, where $L$ is the number of wavelength bins \citep{Flasseur2020PACOSpectrographs}. For the simulated measurements presented above, a spectral resolution of $R=20$ and a wavelength range from \SIrange[]{4}{18.5}{\micro\meter} is used (unless otherwise noted), which results in $L=31$ wavelength bins (with varying bin widths).

\begin{figure}[t]
        \raggedleft
        \includegraphics[width=0.94\linewidth]{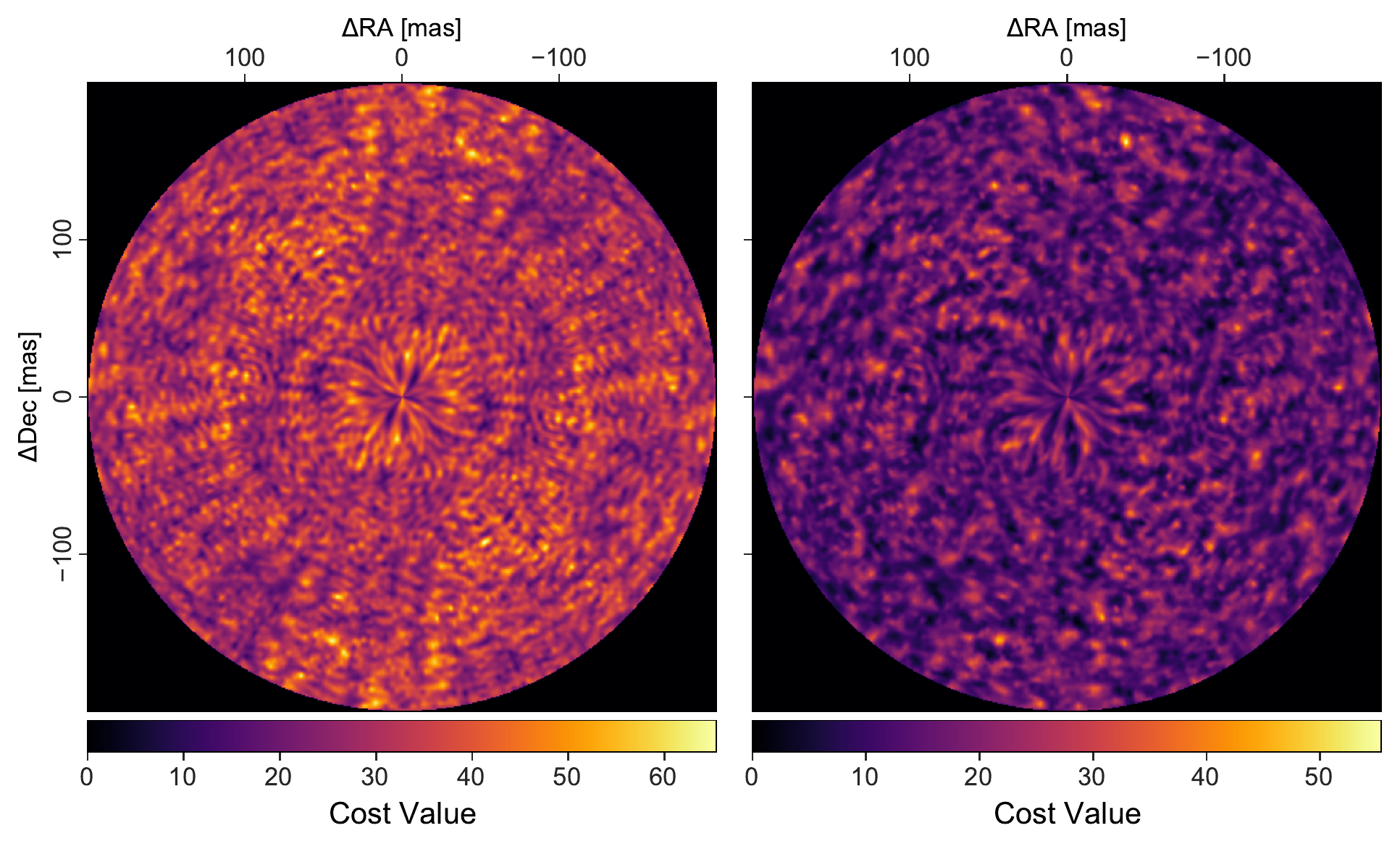}
        \includegraphics[width=1\linewidth]{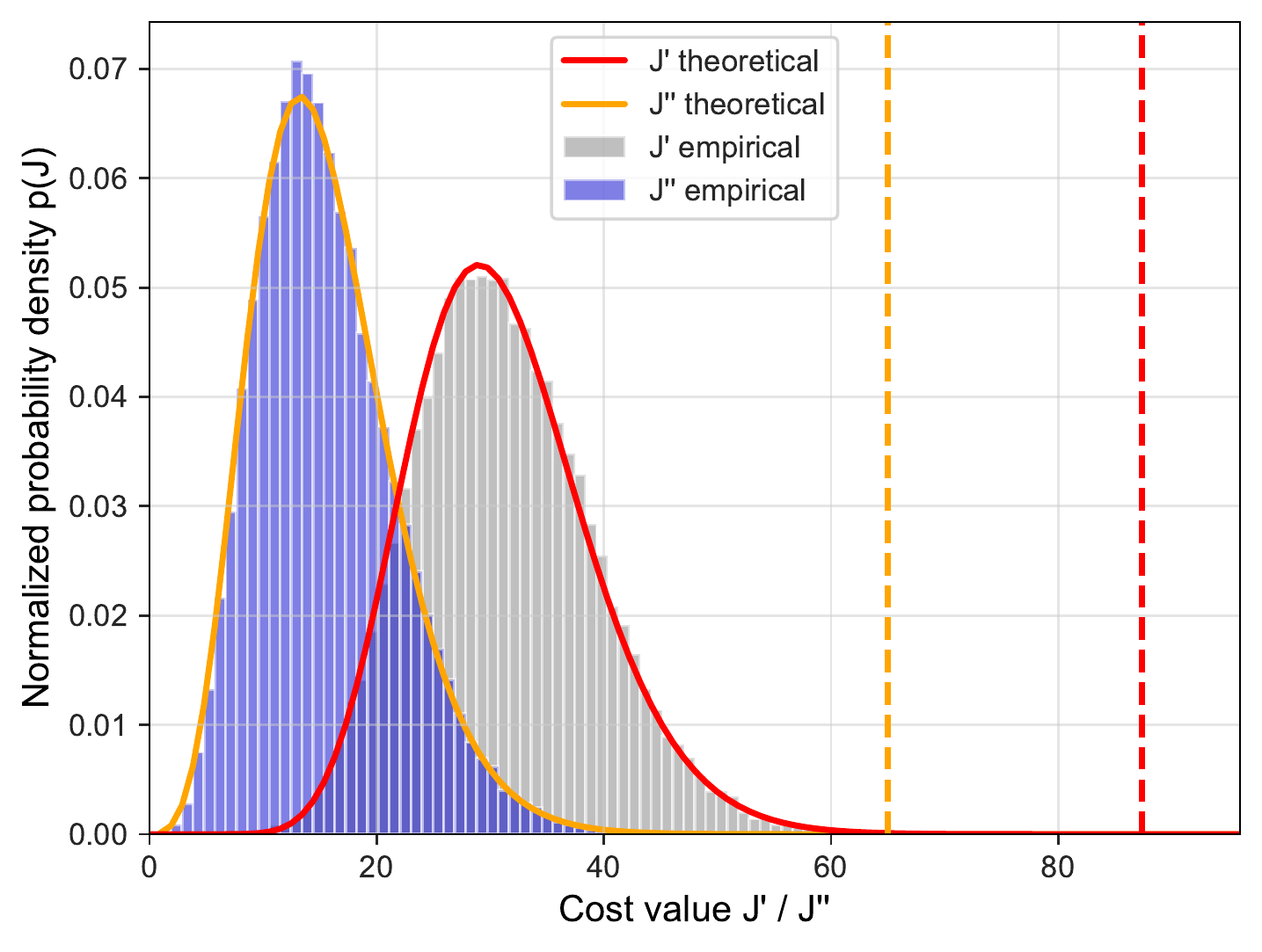}
        \caption{\textit{Top:} Different versions of the cost function $J$ for a simulated observation without any point sources. $L=31$ wavelength bins are used. \textit{Top left}: $J'$ without positivity constraint on the flux. The image is point-symmetric with respect to the image center. \textit{Top right}: $J''$ with a positivity constraint on the planet flux. Compared to the left image, the strict point-symmetry is lifted and the mean value is lower. \textit{Bottom}: Probability distribution of the cost function maps shown above. The grey histogram shows the $J'$ values obtained from the panel on the top-left. The solid red line shows the probability density function of a $\chi^2$-distribution with $L=31$ degrees of freedom. The blue histogram shows the $J''$ values obtained from the panel on the top-right. The solid orange line shows the theoretically expected distribution of the $J''$ values given in Eq.~\eqref{eq:prob_dist_J2}. The dashed vertical lines indicate the detection criteria of $\eta = 87$ and $\eta = 65$ as derived by Eq.~\eqref{eq:eta1_definition} and Eq.~\eqref{eq:J2_detection_criterion} respectively.}
        \label{fig:analysis_cost_func_noplanet}
    \end{figure}

    Figure~\ref{fig:analysis_cost_func_noplanet} (\textit{top-left}) shows the cost function $J'$ calculated for a simulated measurement without any point sources. The non-uniformity of the image is a result of the statistical fluctuations of the background noise.
    The bottom panel in Fig.~\ref{fig:analysis_cost_func_noplanet} shows the corresponding distribution of the $J'$ values in the grey histogram. The simulated data follow the $\chi^2$-distribution with $L=31$ degrees of freedom (red line) very closely. Following the derivation by \cite{Flasseur2020PACOSpectrographs}, the probability of false alarm (PFA) in a data set containing no point sources but only background noise is given by $\mathrm{PFA} = \mathrm{p}(\chi^2_L > \eta)$, where $\eta$ is the detection threshold, which has to be determined. To obtain a PFA corresponding to a 5-$\sigma$ confidence level, one needs to solve
    \begin{align} \label{eq:eta1_definition}
        F_L(\eta) = \dfrac{\gamma(L/2, \eta/2)}{\Gamma(L/2)} = \Phi(5)
    \end{align}
    for $\eta$. Here, $F_L$ is the cumulative distribution function of the $\chi^2_L$ distribution, which can be expressed in terms of the lower incomplete gamma function $\gamma$ and the gamma function $\Gamma$, and $\Phi$ is the cumulative distribution function of the standard normal distribution. 
    Solving  Eq.~\eqref{eq:eta1_definition} for $\eta$ gives a detection threshold for $J'$ of $\eta \approx 87$.
    If the $J'$ value of the most likely planet position is greater than $\eta$ it can be considered a detection.

   Using $J'$ as a detection criterion is, however, unsatisfactory as explained above, and $J''$ - with positivity constraint on the flux - is better suited.  At a single wavelength, and without any point sources being present, $J''_\lambda$ follows a $\chi^2$-distribution with one degree of freedom - the distribution of squares of normally distributed values - with a threshold at 0 due to the positivity constraint on the estimated flux values:
    \begin{align}
        p(J_\lambda'') = \dfrac{1}{2} \delta_0(J''_\lambda) + \dfrac{1}{2} \chi^2_1(J''_\lambda).
        \label{eq:analysis_cost2_stat_dist}
    \end{align}
    The probability distribution of the sum of $L$ independent values following the above distribution is given by
    \begin{align} \label{eq:prob_dist_J2}
        p(J'') = \dfrac{1}{2^L} \delta_0(J'') + \sum_{\lambda=0}^{L-1} \dfrac{L!}{2^L \lambda! (L-\lambda)!} \chi^2_{L- \lambda}(J'').
    \end{align}
    In Fig.~\ref{fig:analysis_cost_func_noplanet} the bottom panel shows as a blue histogram the distribution of the $J''$ values found in the map on the top-right panel. It follows the expected distribution from Eq.~\eqref{eq:analysis_cost2_stat_dist}.
    
    Thus, similar to Eq.~\eqref{eq:eta1_definition}, a detection criterion $J''>\eta$ can be defined by finding $\eta$ such that
    \begin{align}\label{eq:J2_detection_criterion}
        \dfrac{1}{2^L} + \sum_{\lambda=0}^{L-1} \dfrac{L!}{2^L \lambda! (L-\lambda)!} F_{L- \lambda}(\eta) = \Phi(5)
    \end{align}
    with the left hand side being the cumulative distribution function of the probability distribution defined in Eq.~\eqref{eq:prob_dist_J2}.
    
     Solving  Eq.~\eqref{eq:J2_detection_criterion} with $L=31$ for $\eta$ gives a detection threshold of $\eta \approx 65$ for the cost function with positivity constraint. A value of $\eta = 65$ in the cost map $J''$ thus corresponds to a 5-$\sigma$ detection.

    The detection measure presented here does not include spectral regularization. While a regularization term improves the contrast of the detection maps (see, Fig.~\ref{fig:analysis_cost_evolution} in the Appendix), it adds an interdependence between the wavelength bins. Thus, the estimated signal, as well as the noise terms in the individual wavelength bins, are no longer mutually independent. Because the derivation of the detection criterion was based on the assumption of uncorrelated noise, it cannot be applied directly in the case of regularization. However, it was already pointed out by \cite{Mugnier2006DataMission}, and is also indicated by the example presented in Fig.~\ref{fig:analysis_cost_evolution}, that regularization is mostly needed for the detection of planets with a low underlying $\mathrm{SNR}\lesssim 5$.

    \section{Signal analysis}\label{sec:analysis}
\subsection{Single planet: Earth-twin at 10 pc}\label{sec:analysis_earth_twin}

\begin{figure}[h]
        \centering
        \includegraphics[width=0.99\linewidth]{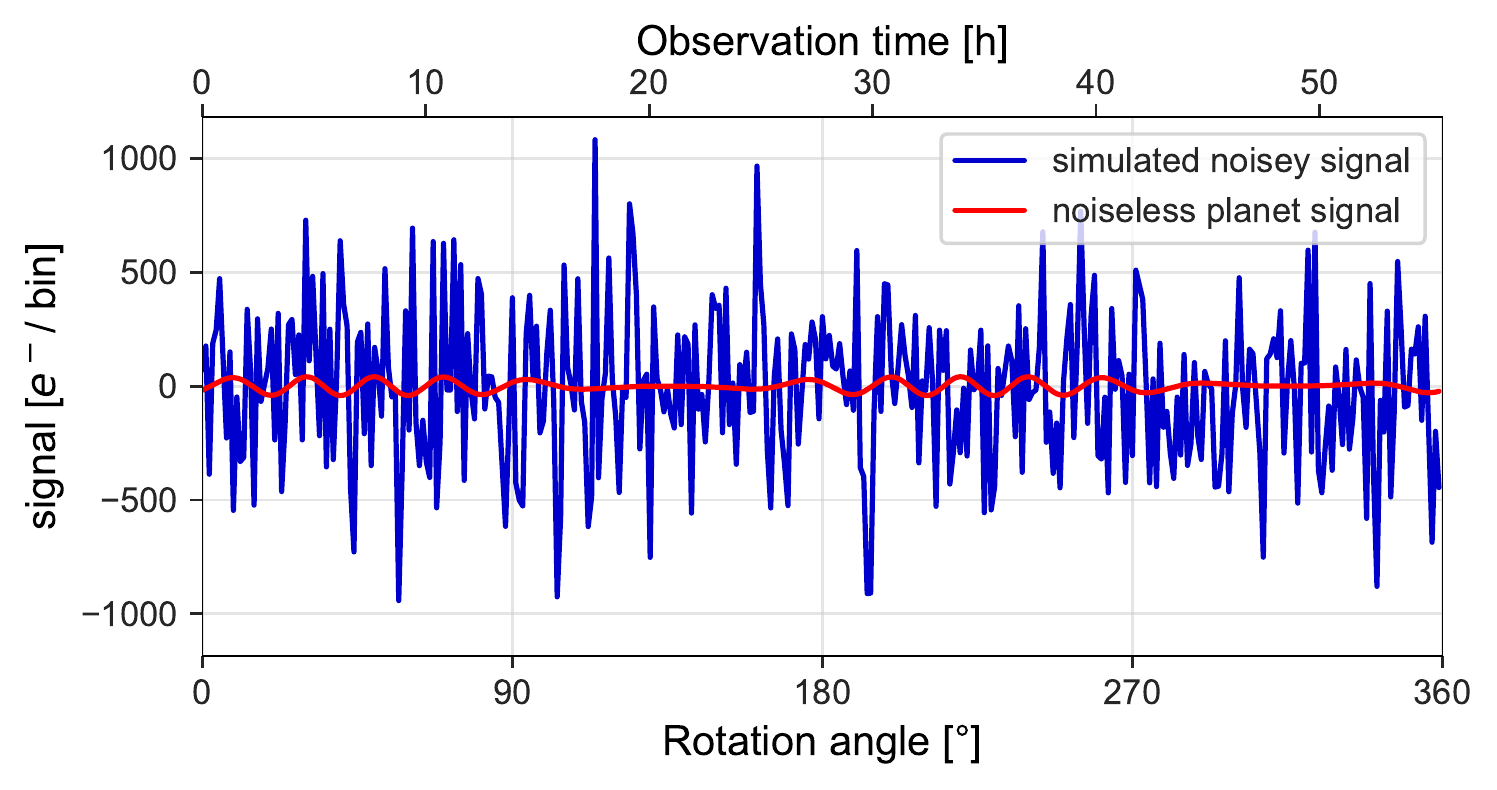}        \includegraphics[width=0.95\linewidth]{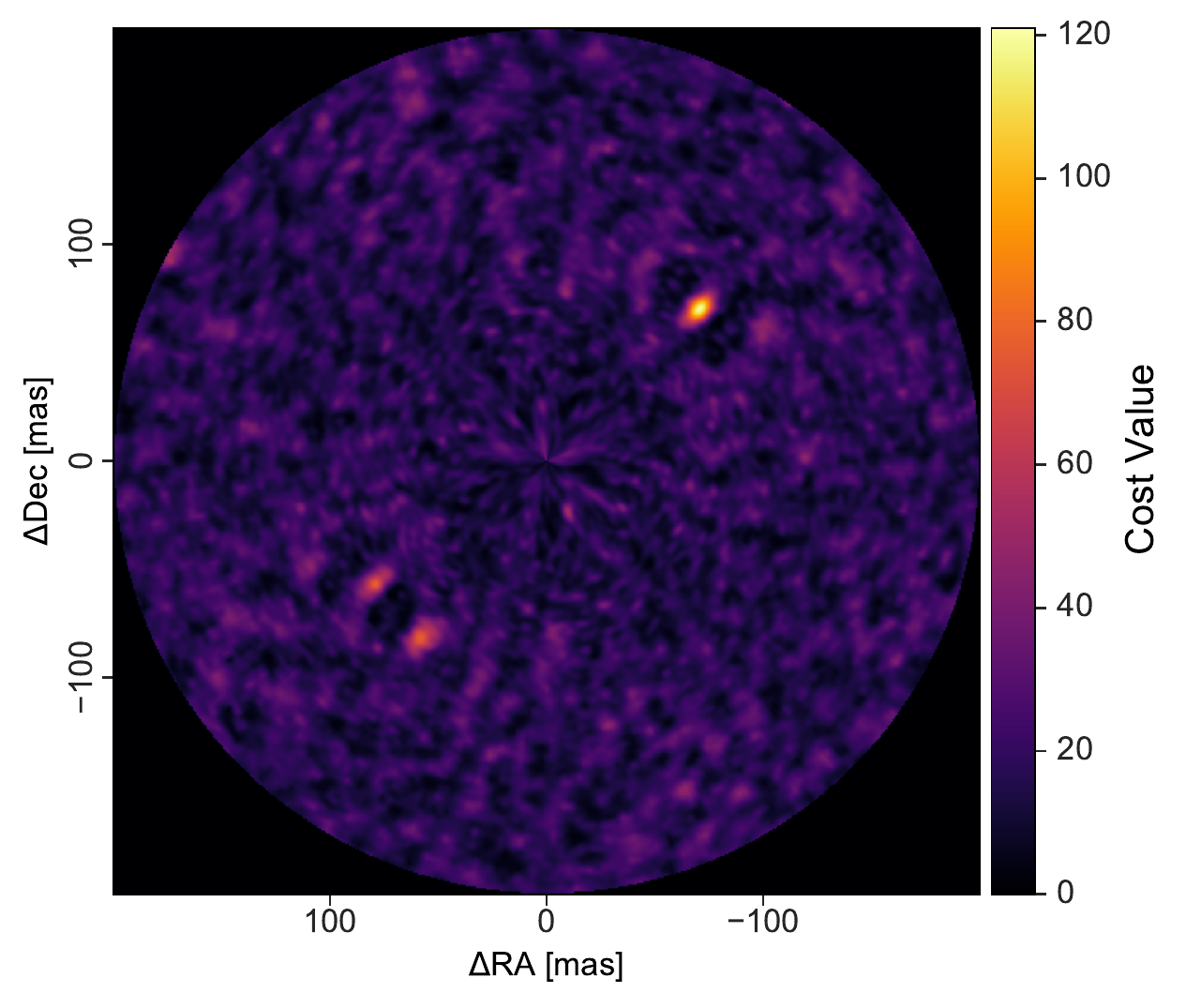}
        \includegraphics[width=0.99\linewidth]{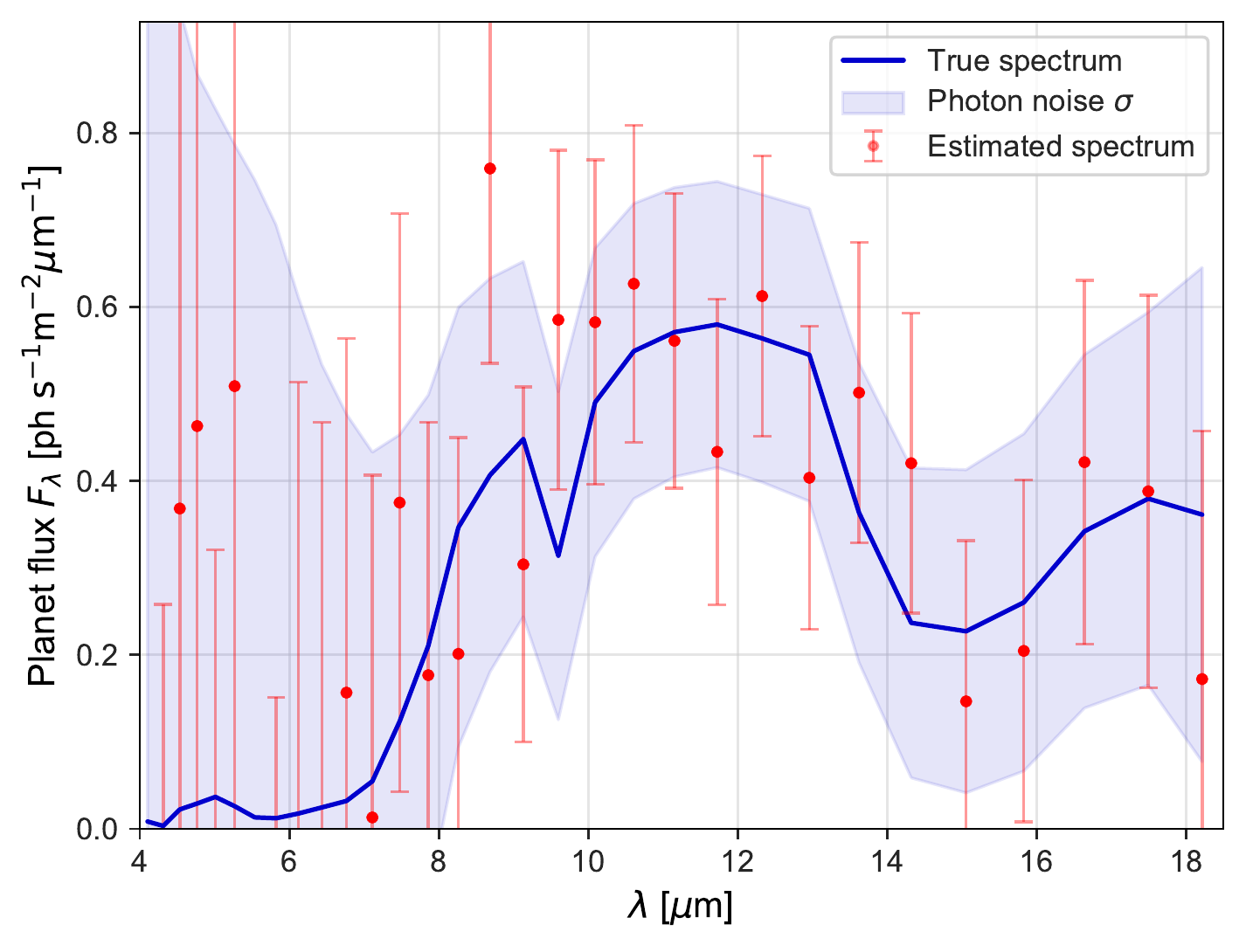}
        \caption{\textit{Top:} Noisy time series signal from simulated data for an Earth-like planet around a Sun-like star at \SI{10}{pc} (in blue). The underlying noiseless planet signal is indicated by the red curve.
        \textit{Middle:} Detection map for simulated data of the Earth-like planet. The photon-noise based SNR is $9.7$ (cf. Sect.~\ref{sec:methods_earth_twin}). The planet is correctly found in the upper right of the map. \textit{Bottom:} Spectral flux of the Earth-like planet as estimated from the simulated data (red data points with $1\sigma$ error bars). The blue line shows the true flux of the simulated planet and the blue shaded region indicates $\pm1\sigma$ based on the photon statistics.}
        \label{fig:analysis_ET_cost_map}
    \end{figure}

    We apply the signal extraction method to the case study example of an Earth-twin planet orbiting a Sun-like star at 10 pc (see, Table~\ref{tab:parameters}). We compare the SNR of the extracted signal to predictions based on photon statistics and determine the position of the signal. We simulate the planet at $\SI{100}{mas}$ projected angular separation from the star. We remind the reader that for an  assumed integration time of \SI{55}{\hour} one obtains a predicted integrated $\mathrm{SNR_{pred}} = 9.7$. The top panel of Fig.~\ref{fig:analysis_ET_cost_map} shows the noisy time series resulting from the simulation.
    The maximum likelihood analysis is performed on a grid out to $\SI{200}{mas}$ angular separation from the central star, with a resolution of $\sim\SI{0.5}{mas}$. The middle panel of Fig.~\ref{fig:analysis_ET_cost_map} shows the cost function map $J''$ without regularization ($\mu=0$) for a single simulated measurement. The relatively high SNR allows for a clear detection of the planet in the upper right corner. The detection criterion as derived in Sect.~\ref{sec:analysis_detection_criterion} is fulfilled and the planet location is estimated correctly within the resolution provided by the grid. The estimated extracted spectral flux of the located planet is shown in the bottom panel in Fig.~\ref{fig:analysis_ET_cost_map}. The extracted spectrum agrees with the underlying true spectrum within the expected uncertainties. 
    
    Estimates for the location and the spectrum have been extracted from the simulated data, but we still need to quantify the SNR and compare it to the expected value. Applying Eq.~\eqref{eq:analysis_snr_int_simple} results in $\mathrm{SNR_{est}} = 10.45$, compared to the predicted $\mathrm{SNR_{pred}} = 9.7$. The discrepancy is due to the statistical nature of the simulated data as mentioned previously. 
    If the simulation is repeated, the outcome is slightly different as would be the case for a real measurement. To better estimate the performance of the signal extraction process in general, and the SNR estimation in particular, the simulated measurement is repeated $1000$ times with the same set-up. For each run the planet position and spectral flux are estimated from the simulated data.
    
\begin{figure}[h]
        \includegraphics[width=1\linewidth]{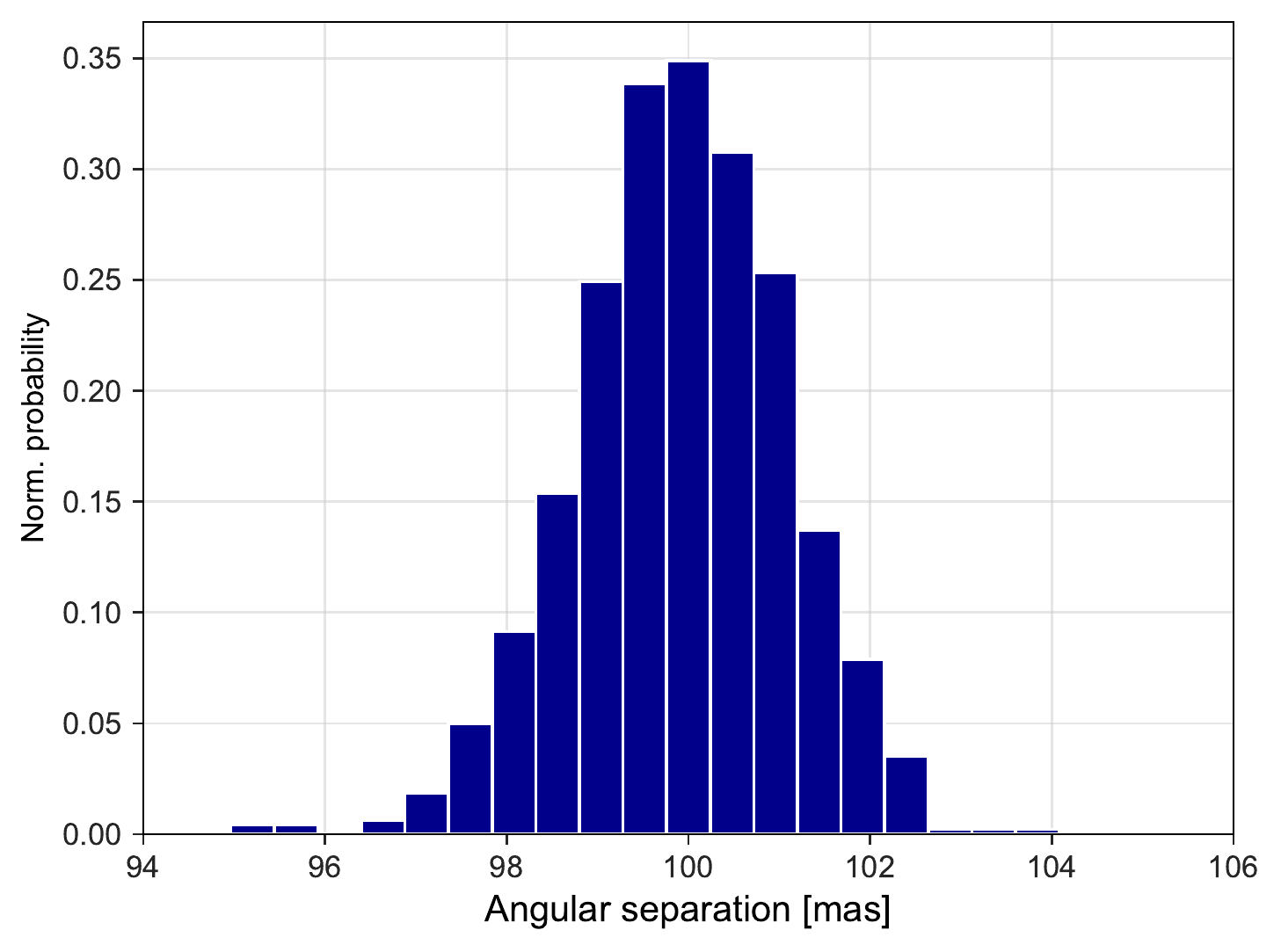}
        \includegraphics[width=1\linewidth]{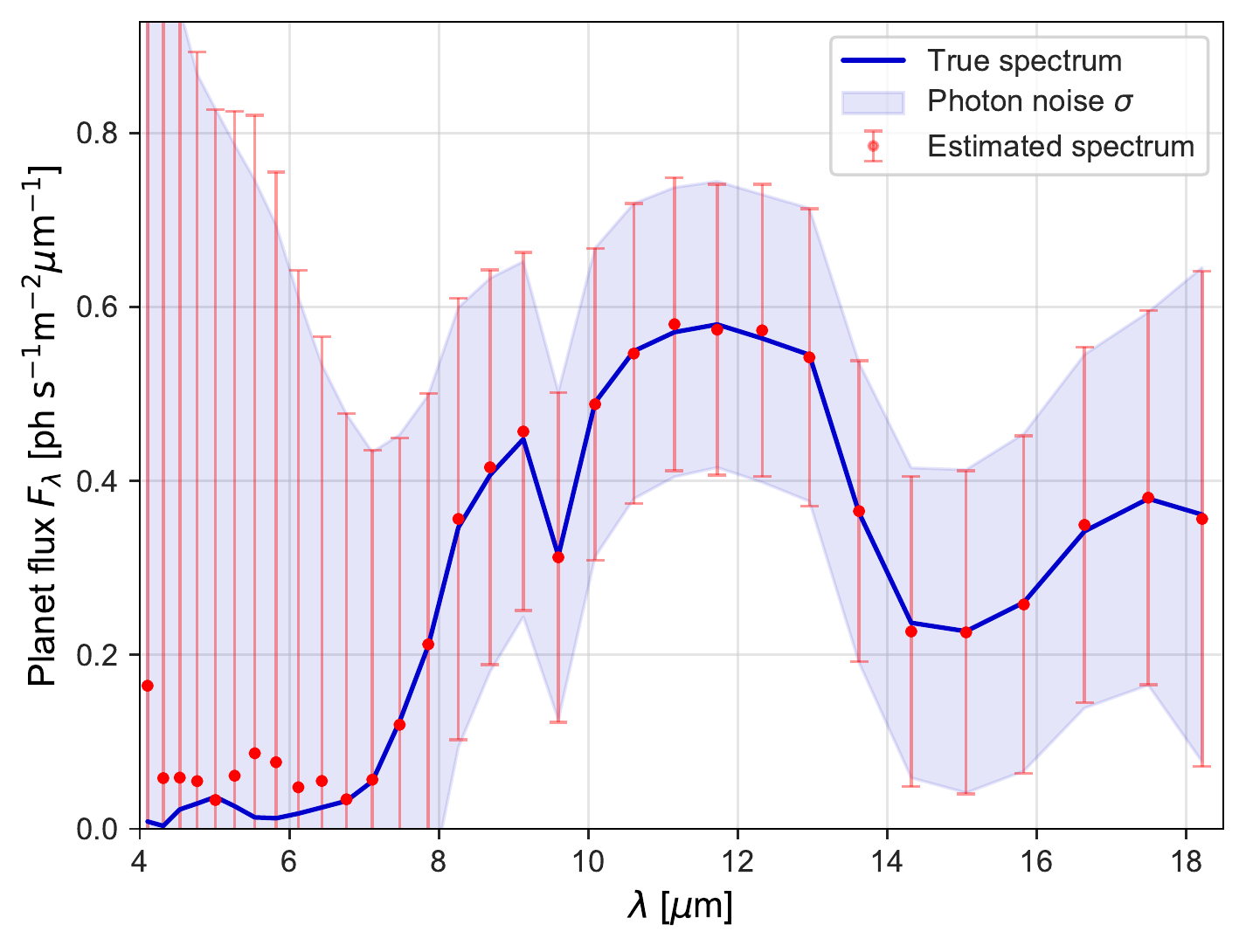}
        \caption{\textit{Top}: Distribution of the estimated angular separation from the host star of 1000 simulated Earth-twins. The underlying true value is $\theta = 100\,\mathrm{mas}$, the average estimated value is $99.7 \pm 1.5\,\mathrm{mas}$. The spacing of the bars corresponds to the angular resolution of the grid on which the analysis was performed. \textit{Bottom}: Mean (red dots) with standard deviation (red error bars) of the distribution of the estimated flux values over 1000 simulations of the same artificial planet. The blue line shows the true spectrum, with the blue shaded region enclosing $\pm1\sigma$ based on the photon statistics.}
        \label{fig:analysis_ET_mc_r_est}
    \end{figure}

    The top panel in Fig.~\ref{fig:analysis_ET_mc_r_est} shows the distribution of the estimated angular separation of the planets. The average of the estimated angular separation is $99.7 \pm 1.5\,\mathrm{mas}$. The azimuthal position was extracted correctly for all simulated planets within the resolution of the grid ($\pm1^\circ$). While the extracted spectra are all noisy and are qualitatively similar to the example shown in the bottom panel of Fig.~\ref{fig:analysis_ET_cost_map}, the bottom panel of Fig.~\ref{fig:analysis_ET_mc_r_est} shows the mean and the standard deviation of the extracted flux values per wavelength bin over the $N=1000$ simulations in comparison with the input spectrum. Apart from the short wavelength range, where the low flux and high noise levels result in large uncertainties, the extracted mean spectrum agrees very well with the input data.

    \begin{figure*}[t]
        \centering
        \includegraphics[width=1\linewidth]{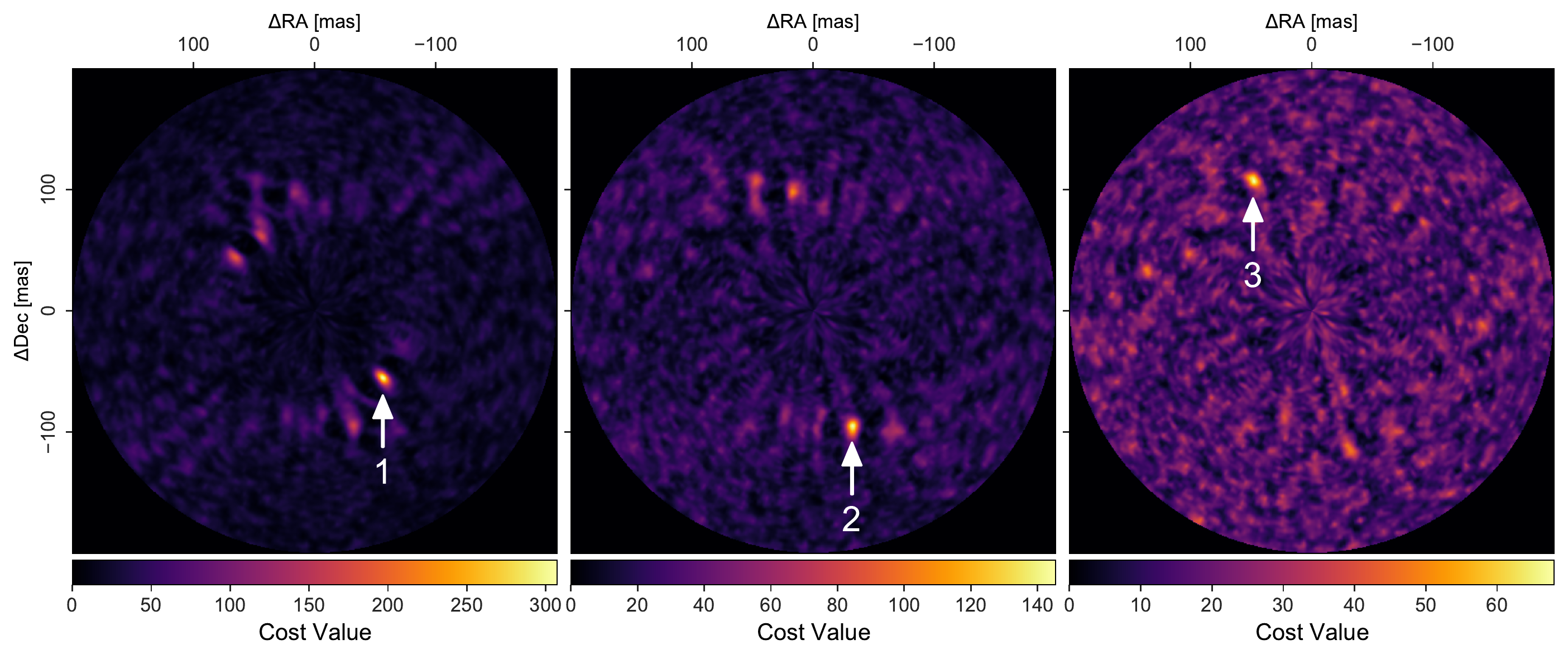}
        \caption{Cost function detection map for the iterative detection of three simulated planets (indicated by the white arrows). The simulation parameters are given in Sect.~\ref{sec:analysis_multi_planet}. The left image shows the detection map calculated from the full simulated data. Subtracting the estimated signal contribution of the source detected in the first map and recalculating the detection map results in the second image, and after repeating in the third one. Removing a detected source from the data also removes the side lobes created by the source, thus allowing for a clear iterative detection.}
        \label{fig:analysis_multiplanet_snr}
    \end{figure*}

\subsection{Multi-planet systems}\label{sec:analysis_multi_planet}
    As most exoplanetary systems are expected to contain multiple planets, it is important that extraction algorithms can deal with multiple objects. The modulation signal recorded with a nulling interferometer is a superposition of the signals from different sources within the field-of-view. 
    The most straightforward approach to differentiate individual planets is to find their positions and the estimated fluxes one after another. To do so, the most likely planet position, in absolute terms, is accepted as a detection provided it fulfills the detection criterion described above, and the spectral flux is estimated for that planet. The estimated planet signal is subtracted from the data and the process is repeated until no other significant signal remains in the data.
    
    We applied the signal extraction process outlined above to a simulated 3-planet system. All planets have a radius $R_\mathrm{p} = 1\,R_\oplus$ and are located around a Sun-like star at \SI{10}{pc}. Their semi-major axes $a_\mathrm{p}$ are set at 0.8, 1.0, and 1.2\,AU respectively. 
    All planets are assumed to radiate as black bodies with an equilibrium temperature of \SI{276}{\kelvin} scaled by the inverse square-root of the semi-major axis, $T_\mathrm{p} \propto 1/\sqrt{a_\mathrm{p}}$. The system is seen face-on, such that all planets appear at their respective maximum angular separation.
    The azimuthal position of the planets is chosen randomly. The system is assumed to harbor an exozodi disk with $z=3$.
    For an assumed integration time of \SI{55}{\hour} the predicted SNRs of the three planets are 15.8, 10.4, and 6.9, respectively.
    
    Figure~\ref{fig:analysis_multiplanet_snr} shows the three detection maps for the iterative signal extraction process applied to the simulated data. In the left panel, the highest values for the cost function indicate the correct position of the planet closest to the host star, which has the highest expected SNR. Besides the side lobes of the first planet, the positions of the other planets, as well as their side lobes, are also visible. However, because the cost values of these features are similar, it is unclear which signal corresponds to a true source and which not. The subtraction of the highest SNR signal removes also its side lobes and the second planet can be identified by the strongest remaining peak. Removing the second signal allows the third source to be detected. After the third source, no further point source is found above the detection criterion. Extracting the signal of the three planets leads to estimated SNR values of 16.0, 10.1, and 7.1,  respectively.
    \rs{To test the robustness of the extraction process with respect to the planet position we iterate the signal extraction over distinct realisations of the noise. We find that the angular separation and the position angle of the planet is retrieved accurately within the given uncertainties. These uncertainties scale in accordance to the respective photon noise SNRs, meaning that the innermost planet is extracted to the highest precision.}

    \begin{figure}[b!]
        \centering
        \includegraphics[width=1\linewidth]{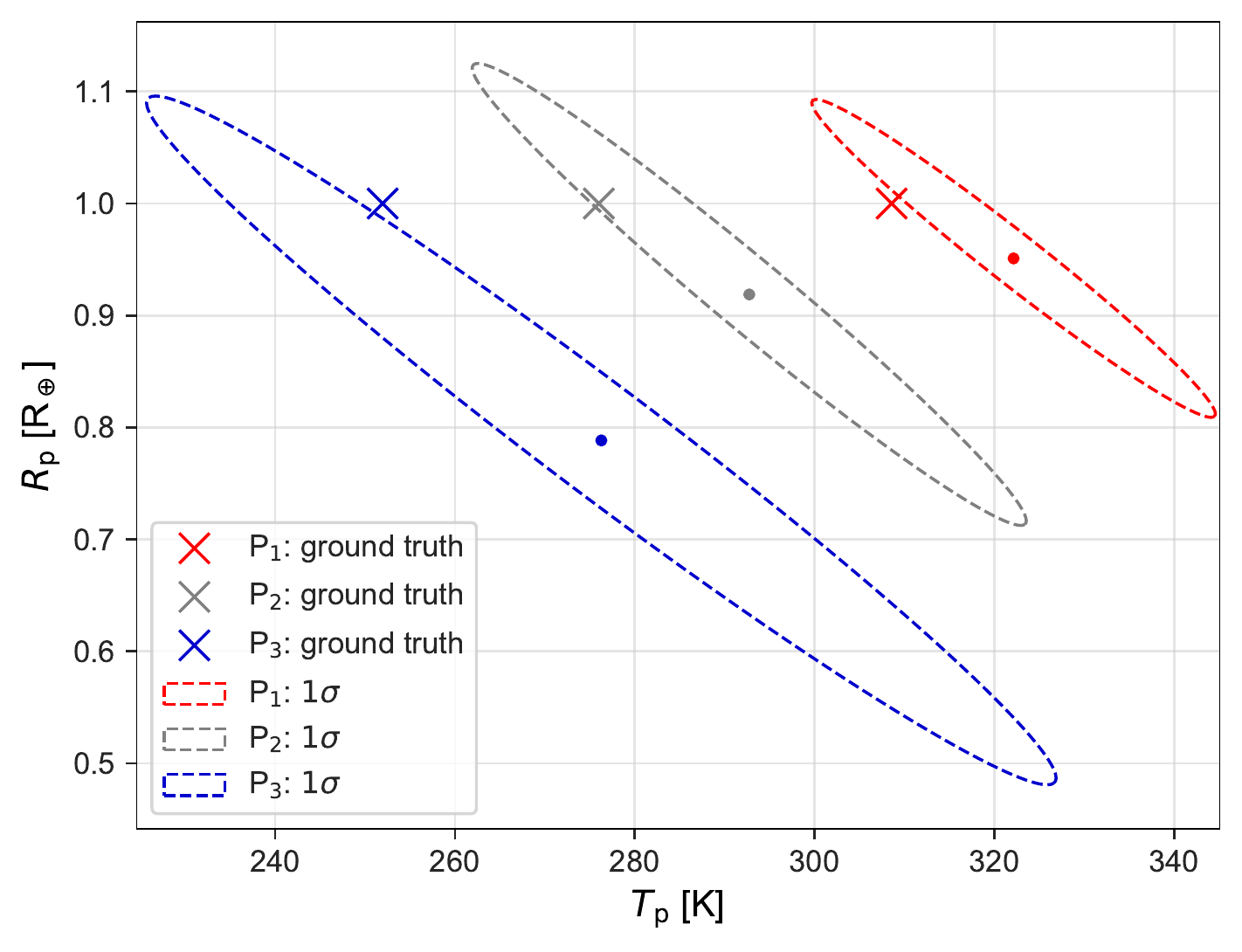}
        \caption{Temperature and radius estimates for the three simulated planets with $R_\mathrm{p}= 1\,R_\oplus$ and varying temperatures. The true values are indicated by crosses, the estimated values by dots with corresponding $1\sigma$ confidence ellipses. The values are estimated by fitting a black body curve to the estimated spectra of the point sources, whose locations are inferred from the detection maps shown in Fig.\,\ref{fig:analysis_multiplanet_snr}.}
        \label{fig:analysis_TR_correlation}
    \end{figure}
    
    We fitted the extracted spectra of the three planets with a black body spectrum to estimate their radii and effective temperatures as this will also be done as one of the first steps for any planet candidates detected with \emph{LIFE}. Figure~\ref{fig:analysis_TR_correlation} shows the estimated values along with $1\sigma$ confidence contours derived from the uncertainties in the extracted flux values. The true values are indicated by crosses. The uncertainty in the estimated temperatures and radii increases with decreasing SNR, from about $\sigma_T \approx$ 20 K and $\sigma_R \approx$ 0.15 R$_\oplus$ for the brightest planet to $\sigma_T \approx$ 35 K and $\sigma_R \approx$ 0.4 R$_\oplus$ for the faintest. As the spectral shape of an emitting black body depends more strongly on the temperature than on its size, the relative uncertainty in the derived temperature is much smaller than the relative uncertainty in the radius. 
    
\subsection{Rocky, habitable zone exoplanets from \emph{LIFE} search phase}
\begin{figure}[t]
        \centering
        \includegraphics[width=1\linewidth]{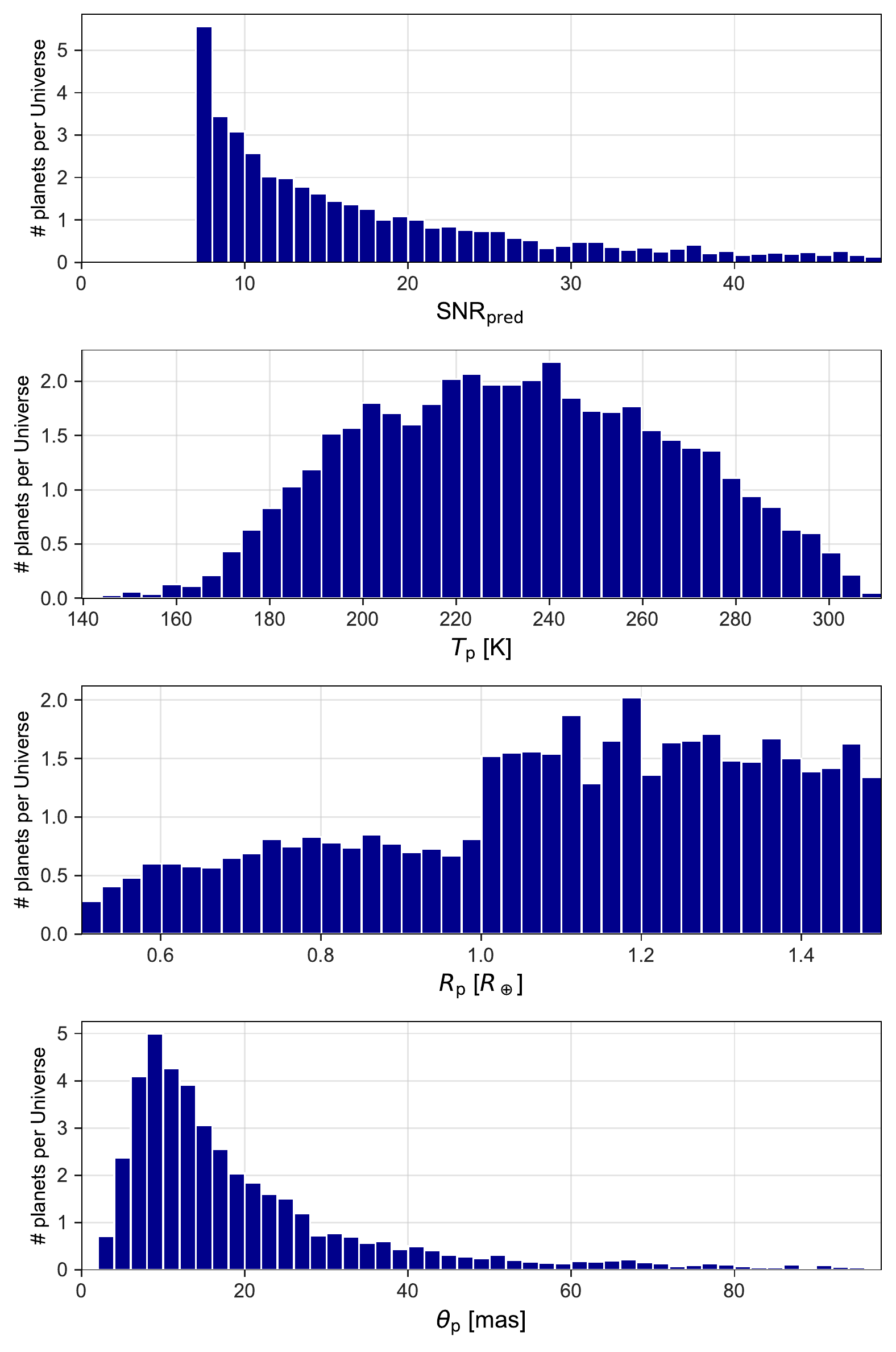}
        \caption{Distribution of properties of detectable rocky, habitable zone planets from the detection yield estimates presented in \cite{paper1} normalized over 100 Monte Carlo runs. From top to bottom: predicted SNR (based on photon-noise), effective temperature, radius, and projected angular separation.}
        \label{fig:analysis_mc_planet_props}
    \end{figure}

    In the previous sections we investigated how well the signal of individual exoplanets can be extracted and their basic properties, such as effective temperature, radius and position, can be estimated. We now turn to a larger sample of objects covering a broader range of properties and different  noise characteristics (e.g., spectral type of host star, exozodi level). The extraction algorithm is applied to a subset of the simulated exoplanet population that was used for the detection yield estimation in Paper 1 \citep{paper1}. Specifically, we focus on the rocky, habitable zone planets\footnote{Defined as planets with radii $R_{\textrm p}$ in the range $0.5\;R_{\oplus} \leq R_{\textrm p}\leq 1.5\;R_{\oplus}$ orbiting within the spectral-type dependent empirical habitable zone (eHZ) of their host star. For a Sun-like star the eHZ corresponds to an insolation range of $0.32\;S_{\oplus} \leq S_{\textrm p}\leq 1.76\;S_{\oplus}$, where $S_{\oplus}$ corresponds to the present-day insolation level of the Earth.} that were detectable in a 2.5-year search phase with four $D = 2$\,m apertures and investigate in detail how well their properties can be determined from single epoch observations, assuming the planets radiate as black bodies at their respective equilibrium temperature. From the 500 Monte Carlo runs that we carried out in \cite{paper1} in order to estimate the expected detection yield of the \emph{LIFE} mission, we randomly select 100 runs. These runs contain approximately 4400 rocky, HZ planets with predicted $\mathrm{SNR}\geq7$, which was considered as the detection threshold in Paper 1. Figure~\ref{fig:analysis_mc_planet_props} shows the distribution of the predicted SNR of these planets (based on photon-noise statistics), their temperature, radius, and their projected angular separation from their host star. The median of the SNR distribution is $\approx$13.3, thus nearly twice the detection threshold. While, on average, the planet radii are $>$1.0$\;R_{\oplus}$, the mean temperature of the detected planets is only $T_\mathrm{p}\,\SI{\approx 235}{\kelvin}$, i.e., less than Earth's effective temperature.
    
    Our signal extraction method is applied to each planet individually and the detection maps are calculated without regularization of the estimated spectra. The detection criterion as derived in Sect.~\ref{sec:analysis_detection_criterion} is applied. We find that $\approx$1.5\% of the simulated planets do not fulfill the detection criterion. For a few additional planets the position was extracted incorrectly. Overall, about 98\% of the planets were detected at a separation deviating $\leq$15\% from the true value and with an error in position angle $\leq$10$^\circ$.
    
\begin{figure*}[ht]
        \centering
        \includegraphics[width=0.9\linewidth]{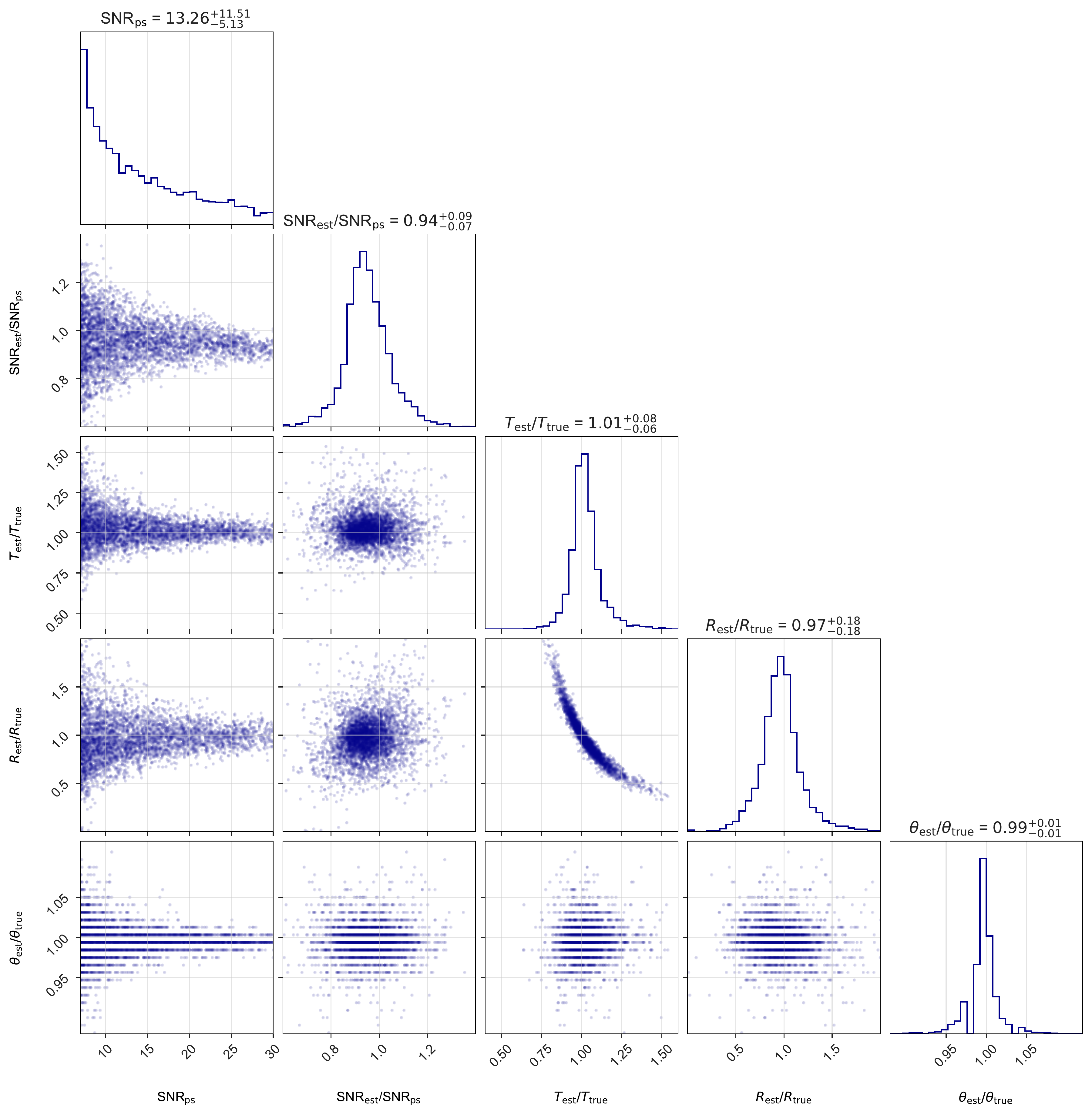}
        \caption{Corner plot showing the predicted SNR based on photon statistics (cf. top panel in  Fig.~\ref{fig:analysis_mc_planet_props}), and the estimated values for SNR, temperature, radius, and angular separation relative to the true values based on the signal  analysis of $\approx4400$ rocky, habitable zone planets. We note that the discretization of the values for the estimated angular separation corresponds to the resolution of the spatial grid on which the analysis was performed. At the top of each column the mean and the standard deviation of the distributions are given.}
        \label{fig:analysis_corner_mc}
    \end{figure*}

    For the detected planets, the temperature and radius are estimated by fitting a black body spectrum to the extracted flux. Additionally, the SNR is estimated from the extracted flux as described in Sect.~\ref{sec:analysis_earth_twin}. Because each of the estimated parameters has a broad distribution among the simulated planets, the relative deviation of the estimated values from the underlying true values is analyzed. 
    The results of the parameter extraction are presented in Fig.~\ref{fig:analysis_corner_mc}.
    The corner plot shows again the distribution of the predicted SNR and the ratios between the estimated and the true values of the planetary parameters (i.e., temperature, radius, and separation) as well as the ratio between the estimated and the purely photon statistic based SNRs.
    From the first column it is evident that the estimation of all parameters improves with increasing SNR of the planet, as the statistical distribution of the estimates becomes narrower. Overall, the SNR is slightly underestimated, even for relatively high SNR values $>25$. This can be explained by an overestimated noise variance from the simulated data, as it is calculated as the variance of the full data including possible planet signals.
    
    The deviation of the estimates of the temperature and the radius from the true values are approximately symmetric in terms of over- or underestimation. Across the full sample, the temperature is estimated to $T_\mathrm{est}/ T_\mathrm{true} = 1.01^{+0.08}_{-0.06}$ and the radius to $R_\mathrm{est}/ R_\mathrm{true} = 0.97\pm0.18$. As expected, the estimate of the temperature and the radius are highly correlated as both parameters are positively correlated with the total emitted radiation of the planet. The separation is estimated to $\theta_\mathrm{est}/ \theta_\mathrm{true} = 0.99\pm{0.01}$.
    
    \section{Discussion and Conclusions}\label{sec:conclusions}

We investigate the signal extraction process for simulated interferometric measurements to study the detectability of potential exoplanets and the ability to characterize them. We implement a maximum-likelihood algorithm previously proposed for the \emph{Darwin} mission and apply it to simulated data. This yields the three important results that (a) this method can be successfully applied to multi-planet systems, (b) the signal extraction method is able to reproduce SNR estimates that are in agreement with expectations from photon-noise statistics \rs{and (c) planetary properties inferred by the signal extraction are of high accuracy. These results are especially interesting in the context of \cite{paper1}, since this publication predicts the exoplanet yield of the LIFE mission based on a synthetic planet population without considering the impact of signal extraction. Our results indicate that an inclusion of the signal extraction in these simulations would not fundamentally alter the results.}

\rs{Yet, under certain conditions the signal extraction is especially demanding.} In the low-SNR regime, we have qualitatively shown how spectral regularization can help to disentangle the planet signal from the background noise. However, as spectral regularization relies on a priori assumptions about the planet spectrum, it is not free of bias and adds a degree of freedom to the analysis. To apply regularization to a large set of simulated planets, a detection criterion depending on the amount of enforced spectral regularization would have to be defined. Additionally it has to be investigated how the regularization affects the estimated planet spectrum quantitatively. 

To further assess potential limitations when dealing with multi-planet systems, studies with more diverse planet properties or very small projected angular separations could be envisaged. Potential improvements when dealing with several signals were already presented by \cite{Thiebaut2005MaximumInterferometer}. These authors discussed an extrapolation of the maximum-likelihood method to estimate the planet positions and spectra for multiple planets at once, which avoids error propagation from the estimation of the first detected planet to the other ones. This generalized approach is computationally more expensive, but a simplified option  estimates the planet positions one after the other but recomputes the optimal flux of all detected planets at each iteration.

\rs{Furthermore, the ability to correctly extract multi-planet systems is also governed by the effective angular resolution of the array. Therefore, a study of diverse multi-planet extractions would be able to provide minimum requirements for the angular resolution. Since the angular resolution has been shown to be a dominant factor in the design of nulling interferometers \citep{Lay2005ImagingInterferometers}, such a study will become vital in the ongoing development of \emph{LIFE}.}

Going forward, one could easily imagine that new approaches involving modern machine-learning based signal extraction algorithms should be tested and their performance compared. While already being explored in other fields \citep[e.g.,][]{Cuoco2020,Gebhard2019,Gebhard2020}, nulling interferometry was not yet a focus point of these efforts. This will be particularly important as soon as systematic noise terms can be simulated in a realistic way.

\fis{An additional factor degrading the quality of the measurement can be the occurrence of instrumental noise. Within the \emph{LIFE} project, the current working assumption is that the instrumental noise can be neglected based on its contribution being significantly smaller than the fundamental noise. We show, that this assumption places strict requirements on the instrumental stability and prompts continued technical developments, for example in the fields of low dark-current detectors or highly precise OPD control techniques.}

\rs{Similarly, signal extraction has been shown to play an integral role in the prediction of instrumental noise \citep{Lay2004SystematicInterferometers}. Certain steps in the signal retrieval, for example the explicit fitting and removal of systematic noise, can significantly relax technical requirements \citep{Lay2006}. As \textsc{LIFEsim} progresses to include the simulation of instrumental noise, the removal of the same should be considered with the methods presented here.}

\fis{In future work, we aim to shift from the fundamental noise limited regime into a regime in which some instrumental noise is retained in the signal. A prerequisite for this is that the instrumental noise is well understood, tracked and considered in the interpretation of the measurements. \cite{Defrere2010NullingMissions} offers a list of techniques targeted at facilitating this treatment of instrumental noise. The goal of this future work will be to achieve the scientific output postulated by the \emph{LIFE} mission with technology closer to contemporary and near-future instrumentation.}

\rs{In summary, we present a publicly available\footnote{\url{www.life-space-mission.com}} tool for simulating the exoplanet search phase of the \emph{LIFE} mission called \textsc{LIFEsim}\footnote{Because the current version of \textsc{LIFEsim} has already been applied to a number of science case simulations, we would like to enable the community to verify and reproduce the results, and also to investigate additional science cases.}. Using the presented planetary signal extraction methods, we show that the synthetic exoplanet population used for the mission yield evaluation in \cite{paper1} can be retrieved. We demonstrate the precise extraction of planetary radius and temperature, which is of special interest since they are fundamentally important parameters for characterizing the (atmospheric) properties of exoplanets and for categorizing and prioritizing objects for potential in-depth follow-up investigations. We remind the reader that these two parameters are much more difficult to derive from reflected light measurements as there is a degeneracy between planet size and geometric albedo and no immediate information about the effective temperature can be obtained. These results illustrate the high quality of information one can expect from single-epoch observations with \emph{LIFE} and how they will complement the results from future reflected light missions.}

    
    \begin{acknowledgements}
    \rs{We thank the anonymous referee for a critical and constructive review of the original manuscript which helped improve the quality of the paper significantly.} This work has been carried out within the framework of the National Centre of Competence in Research PlanetS supported by the Swiss National Science Foundation. SPQ acknowledges the financial support of the SNSF. TL was supported by the Simons Foundation (SCOL award No. 611576). This research has made use of the following Python packages: 
    \texttt{astropy} \citep{Astropy_2013, Astropy_2018},
    \texttt{matplotlib} \citep{Hunter_2007}, 
    \texttt{numpy} \citep{VanDerWalt_2011},
    \texttt{scipy} \citep{Virtanen_2020}.\newline\newline
    
    \emph{Author contributions:} MO and FD contributed equally to this paper. MO wrote the original \textsc{LIFEsim} tool, carried out the main analyses and wrote part of the manuscript. FD contributed to the main analyses and the \textsc{LIFEsim} tool, programmed the publicly available github version, led the instrumental noise analyses and wrote part of the manuscript. SPQ initiated and guided this project and wrote part of the manuscript. RL, EF and AGh contributed to the \textsc{LIFEsim} tool. All authors discussed the results and commented on the manuscript.
    \end{acknowledgements}


    \bibliographystyle{aa}
    \bibliography{main.bib}
    
    \appendix
    
    \section{Fundamental noise limit} \label{app:fundamental_noise}
Section \ref{sec:instrumental_noise} states that for the analysis presented in this paper a measurement dominated by photon noise originating from the astrophysical sources is assumed. In the following, this state will be referred to as the fundamental noise limited regime. Considering the technical complexity of nulling interferometers, it is important to substantiate this assumption and offer proof that it is well founded. The aim of this appendix is the implementation of an instrumental noise model and the subsequent prediction of the maximum levels of perturbations to the instrument that allow the system to stay in the fundamental noise dominated regime.

\subsection{Assumptions} \label{app:fundamental_noise_assumptions}
The term fundamental noise limited is readily used in literature as well as in the \emph{LIFE} paper series. Under the goal of producing quantitative results, this must be translated into a more tangible constraint. 
The perfect instrument will only be subject to the photon noise by the astrophysical noise sources $\sigma_\mathrm{fund}$\footnote{Square-root of the number of photons contributing to photon noise per unit time}. Perturbing the instrument will give rise to two additional types of noise terms. First and foremost, a systematic perturbation (e.g., a variation in the amplitude response of one of the interferometric arms) of the instrument will introduce a systematic noise term $\sigma_{s, \mathrm{inst}}$ arising from an additional photon rate introduced by the perturbations. These additional photons can resemble a signal as would be produced by a target exoplanet. Second, additional photon noise sources like a detector dark current or thermal emission within the instrument can be considered. Moreover, the systematic perturbations of the central destructive fringe will average to a mean reduction in the effective null-depth. This will increase the overall amount of photons received from the target star and subsequently also increase the photon noise. We collect all additional photon noise sources into the instrumental photon noise term $\sigma_{p, \mathrm{inst}}$ and define the instrumental noise term
\begin{align*}
    \sigma_\mathrm{inst} = \sqrt{\sigma_{s, \mathrm{inst}}^2 + \sigma_{p, \mathrm{inst}}^2}.
\end{align*}
The fundamental noise limited case is then interpreted as the measurement being dominated by the photon noise arising in the perfect instrument over the instrumental noise. Therefore, we define the fundamental noise limited case as
\begin{align}
\sigma_\mathrm{fund} \geq \sigma_\mathrm{inst}, \label{eq:pn_limit}
\end{align}
where the fundamental noise contribution is larger that the instrumental noise contribution. Semantically, it could also be argued that the fundamental noise being a multiple of the instrumental noise could be required. Therefore, we also present the case of a five-times larger fundamental noise in the following.


For the array and instrument configuration, we mirror the setup described in Section \ref{sec:nulling} and Table \ref{tab:parameters}. This includes the simulation of phase-chopping, which is assumed to be facilitated by subtracting two simultaneous realisations of phase-inverted outputs. This simplifies the calculation of instrumental noise terms as, contrary to \cite{Lay2004SystematicInterferometers}, we do not have to include errors at the chopping frequency. All astrophysical sources are simulated according to the models described in Section \ref{sec:astro_sources}. If not stated otherwise, this appendix analyses an observation of an Earth-twin around a Sun-like star located at 10 pc distance with a $z=3$ exozodi level (cf. Table \ref{tab:parameters}). \\



In the following simulations, we assume a non-perfect instrument affected by perturbations. Systematic perturbations are assumed to affect the amplitude response $A$, the phase response $\phi$ and the polarization rotation $\theta$ of each of the four interferometric arms as well as the position $x$ and $y$ of the collector spacecraft. \\
These dynamic perturbations follow a perturbation spectrum centered around values for the configuration of the ideal instrument\footnote{We note the lack of DC-components in the perturbation spectra}. In this work, we omit any constant static offsets from the ideal instrument configuration (e.g. a constant amplitude response offset in  one of the arms). This omission is based on static offsets being more easily calibrated and removed compared to dynamic perturbations.
Due to a lack of reliable empirical data, we stay agnostic to how these perturbations are generated and assume the general noise profiles listed in Table \ref{tab:noise}.
Apart from the aforementioned general systematic perturbations, we examine two more sources of perturbation in an effort to illustrate the allowed magnitude of the remaining perturbation terms. We consider the detector dark current $I_D$ and the thermal background seen by a detector in an environment of temperature $T$. We assume a detector with physical pixel dimensions of the JWST/MIRI detectors \fis{\citep[\SI{25}{\micro\meter} pixel pitch, see][]{rieke_mid-infrared_2015}} and a spectral sampling of 2.2 pixels per wavelength channel. 

\begin{table*}[ht]
\centering
\caption{Parameters of the instrument perturbations at a reference wavelength of \SI{10}{\micro\meter}.}
\begin{tabular}{@{}llllc@{}}
\hline \hline
Perturbation                              & Shape             & Cutoff   & rms       & Wavelength Dependence \\ \hline
Amplitude $\delta A$                      & pink noise, no DC & 10 kHz   & 0.1 \%    & $\lambda^{-1.5}$      \\
Phase $\delta \phi$                       & pink noise, no DC & 10 kHz   & 0.001 rad & $\lambda^{-1}$        \\
Polarization $\delta \theta$              & pink noise, no DC & 10 kHz   & 0.001 rad\tablefootmark{a} & none                  \\
Collector Position $\delta x$, $\delta y$ & white noise       & 0.64 mHz & 1 cm      & none                  \\ \hline
\end{tabular}
\tablefoot{Values taken from \cite{Lay2004SystematicInterferometers}. \tablefoottext{a}{Decreased by an order of magnitude due to a typing error in the original publication (Dèfrere 2021, private communication)}}
\label{tab:noise}
\end{table*}

\subsection{Methods}
The simulations are largely based on the approach presented in \cite{Lay2004SystematicInterferometers}. This paper was published in the context of the TPF-I mission and offers a comprehensive framework for estimating the instrumental noise for nulling interferometers. The following paragraphs will sparsely trace the calculations performed in \cite{Lay2004SystematicInterferometers} to highlight fundamental properties.

First, the overall interferometric response is decomposed into a sum of responses of the individual baselines
\begin{align}
    N = \sum_j \sum_k A_j A_k \left[\cos(\phi_j-\phi_k)\overline{B}_{\mathrm{sym}, jk} - i \sin(\phi_j-\phi_k)\overline{B}_{\mathrm{asym}, jk}  \right]. \label{eq:photon_rate}
\end{align}
Here, $N$ is the detected photon rate, $A_n$ and $\phi_n$ are the amplitude and phase responses of the $n$ interferometric arms and $\overline{B}_{\mathrm{(a)sym}, nm}$ is the Fourier transform of the (anti-) symmetric sky brightness distribution under the baseline from the $n^\mathrm{th}$ to the $m^\mathrm{th}$ collector spacecraft. Equation \eqref{eq:photon_rate} reflects an intrinsic symmetry property of nulling interferometers: baselines with $\pi$-multiple phase difference respond to symmetric (with respect to the line of sight) source while baselines with $\pi/2$-multiple phase differences respond to anti-symmetric sources.
The sensitivity of the photon rate against perturbations of the instrument (here in phase $\delta \phi_n$ and relative amplitude $\delta a_n = \delta A_n / A_n$) is captured in a second order Taylor expansion
\begin{align}
    \delta N \approx & \sum_j \left[ \frac{\partial N}{\partial A_j} A_j \delta a_j 
    +\frac{\partial N}{\partial \phi_j} \delta \phi_j \right] \nonumber \\ 
    & + \sum_j \sum_k \left[ \frac{1}{2} \frac{\partial^2N}{\partial A_j \partial A_k} A_j A_k \delta a_j \delta a_k \right. \nonumber\\
    & \left. + \frac{\partial^2N}{\partial A_j \partial \phi_k} A_j \delta a_j \delta \phi_k 
    + \frac{1}{2} \frac{\partial^2N}{\partial \phi_j \partial \phi_k} \delta \phi_j \delta \phi_k \right]. \label{eq:taylor}
\end{align}
The partial derivatives in this equation represent the sensitivity that the photon rate exhibits against a perturbation of respective kind and order. We note the appearance of second order terms as well as amplitude-phase cross terms. \\
At this point in the calculation, the symmetry of the sources is used to simplify the expressions and arrive at the sensitivity to the perturbations. This will be demonstrated using the sensitivity to stellar leakage. In first approximation, the brightness distribution of the stellar disk is centrally symmetric, leading to $\overline{B}_{\mathrm{*, asym}, nm} = 0$. This reduces Equation \eqref{eq:photon_rate} to 
\begin{align*}
    N_* = \sum_j \sum_k A_j A_k \cos(\phi_j-\phi_k)\overline{B}_{\mathrm{*}, jk}.
\end{align*}
Now the sensitivity of the photon rate towards, e.g., first order amplitude perturbations is given by
\begin{align*}
    C_{A_j}^* = A_j \frac{\partial N_*}{\partial A_j} = 2 A_j \sum_k A_k \cos(\phi_j-\phi_k)\overline{B}_{\mathrm{*}, jk}.
\end{align*}
$C_{A_j}^*$ is called the first order stellar leakage sensitivity coefficient and all remaining sensitivity coefficients in Equation \eqref{eq:taylor} can be determined in the same fashion. \\
In a final step, the photon rates $N$ and $\delta N$ are cross-correlated with planetary template functions, which represents the signal extraction process. The photon noise of the instrument then corresponds to the square-root of the mean photon rate and is therefore given by
\begin{align*}
    \sigma_\mathrm{p} = \sqrt{N + \langle\delta N\rangle},
\end{align*}
where $\langle \rangle$ denotes the ensemble average over all perturbed states of the system. In this description, we identify the following correspondences to the variables defined in the previous Section \ref{app:fundamental_noise_assumptions}: $N = \sigma_\mathrm{fund}^2 t_\mathrm{int}$ and $\langle\delta N\rangle = \sigma_\mathrm{p, inst}^2 t_\mathrm{int}$ where $t_\mathrm{int}$ is the integration time. \\
The systematic noise is accessed via 
\begin{align*}
    \sigma_\mathrm{s, inst} = \sqrt{\langle \delta N^2  \rangle}.
\end{align*}
Up to second order, the method of phase-chopping is able to remove all systematic noise contributions except for one  first order phase ($\delta \phi_n$) and one amplitude-phase cross term ($\delta a_n \delta \phi_m$). We remind the reader that the order of the term is not connected to it's impact but refers to the proportionality between the perturbation and the photon rate in Equation \eqref{eq:taylor}.
Identifying the perturbation with the noise spectra given in Table \ref{tab:noise} enables the calculation of the required noise sources. We refer the reader to \cite{Lay2004SystematicInterferometers} for the full details of the methods described in this section.

\subsection{Results}
\begin{figure*}[h]
    \centering
    \includegraphics[width=0.8\linewidth]{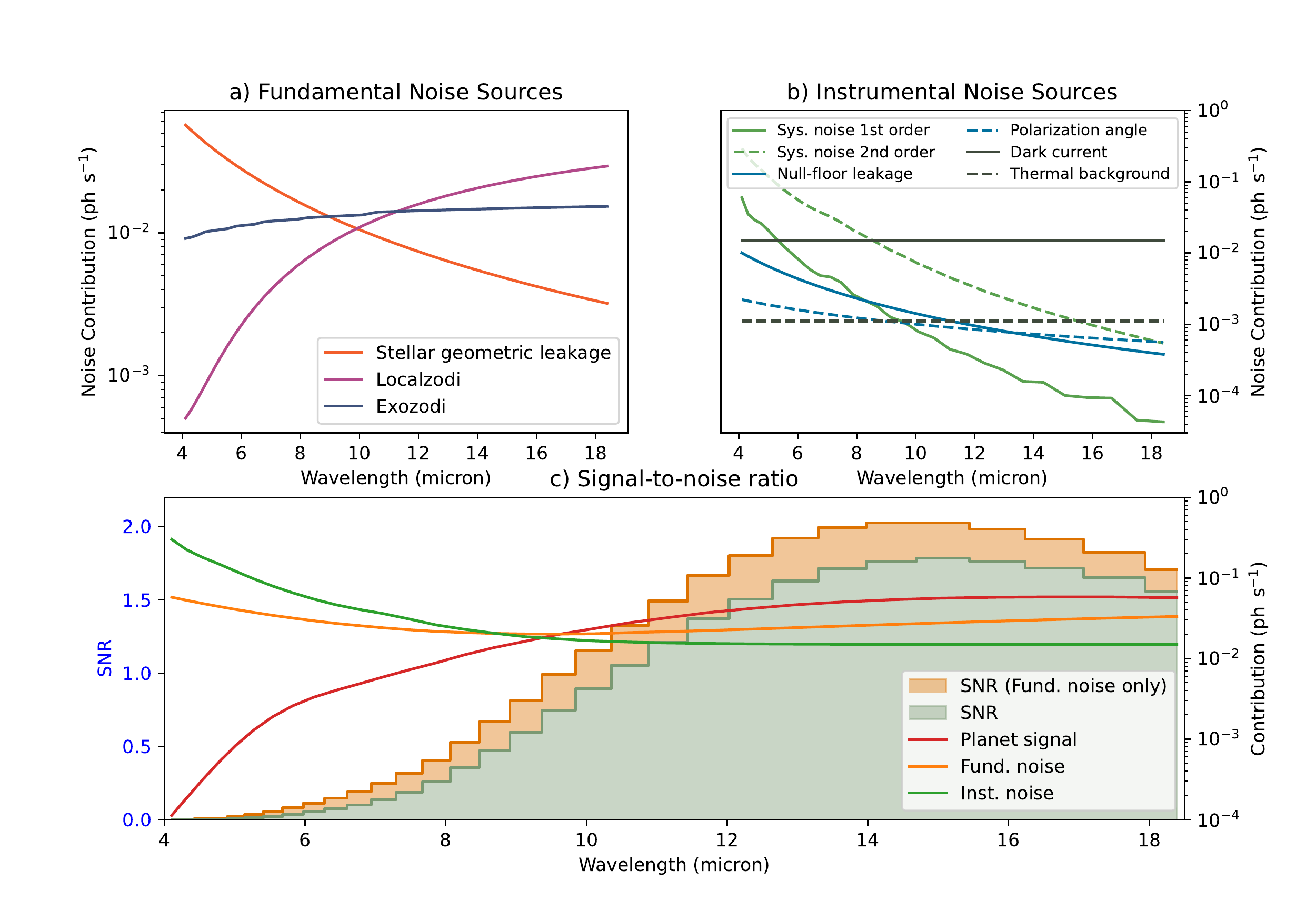}
    \caption{Wavelength dependence of noise contributions for observation of an Earth-twin located at 10 pc with \emph{LIFE} in the baseline scenario. \textit{Panel a):} Division of the fundamental noise into its individual sources. \textit{Panel b):} Instrumental noise sources split into systematic noise sources (green) and additional photon noise sources arising from instrumental effects. \textit{Panel c):} Planet signal, fundamental noise and instrumental noise contributions. A SNR using only fundamental noise (orange bars) is representing the fundamental noise limited case used in the \emph{LIFE} paper series. The green bars show the full SNR additionally considering the instrumental noise as well as the fundamental noise.}
    \label{fig:wavelength2m}
\end{figure*}
The signal and noise contributions using the assumptions laid out above for every wavelength channel defined in the \emph{LIFE} baseline scenario are shown in Figure \ref{fig:wavelength2m}. For the comparison of the fundamental noise sources in Panel a), we reconfirm the behaviour described in Section \ref{sec:methods_earth_twin}: The short-wavelength end of the spectrum is dominated by the stellar leakage term while the long-wavelength end of the spectrum is dominated by the local-zodiacal dust emission, which can not be suppressed by nulling due to its incoherent nature.
For the first and second order systematic noise terms shown in Panel b) we stress again that the use of phase chopping removes all systematic noise terms except for the first order phase deviations and the second order amplitude-phase deviation cross terms. This removal helps to relax the technical requirements imposed by the systematic noise. We reproduce the finding of \cite{Lay2004SystematicInterferometers} that the second order term is generally larger than the first order term. Panel b) also shows that the instrumental noise is either white or increases towards shorter wavelengths. This is driven by the perturbation scaling with wavelength described in Table \ref{tab:noise} and by the increased amount of stellar light emitted in this regime. \\
Panel c) of Figure \ref{fig:wavelength2m} presents the impact of the additional noise sources on the SNR. Here, we assume a detector environment temperature of $T = \SI{11}{\kelvin}$ and a dark current of $I_\mathrm{D} = 10^{-4} \, \, \mathrm{e^- \, px^{-1} \, s^{-1}}$. As expected, the largest relative SNR reduction is induced towards the shorter wavelength end of the spectrum where the systematic noise is most dominant. \\
In the fundamental noise limit, the condition in Equation \eqref{eq:pn_limit} must be valid for all wavelength bins. The relative trends of instrumental and fundamental noise in Figure \ref{fig:wavelength2m} indicate that this condition should be tested towards the shorter wavelength end of the spectrum. We therefore select three discrete wavelength bins below \SI{10}{\micro\meter} for evaluation\footnote{Bins at \SI{4.1}{\micro\meter}, \SI{6.1}{\micro\meter} and \SI{10.1}{\micro\meter},  with widths $\Delta \lambda_b = \SI{0.2}{\micro\meter}$, $\Delta \lambda_b = \SI{0.3}{\micro\meter}$ and $\Delta \lambda_b = \SI{0.5}{\micro\meter}$}.
For completeness, we list the numerical values for the relative signal and noise contribution in this shortest wavelength bin in Table \ref{tab:4micron}. 

\begin{table}[h]
\centering
\caption{Values for the signal and noise contributions for the \emph{LIFE} baseline case with perturbations as defined in Table \ref{tab:noise} for the shortest wavelength bin of $\lambda_b = \SI{4.1}{\micro\meter}$.}
\begin{tabular}{@{} lr @{}}
\hline \hline
Source                       & Photon Rate ($\mathrm{s^{-1}}$) \\ \hline
Planet Signal                & $1.1 \cdot 10^{-4}$           \\ \hline
Fundamental Noise            & $5.8\cdot10^{-2}$            \\
\hspace{1mm} Stellar geometric leakage    & $5.7\cdot10^{-2}$       \\
\hspace{1mm} Local zodi leakage           & $5.0\cdot10^{-4}$           \\
\hspace{1mm} Exo zodi leakage             & $9.1\cdot10^{-3}$           \\\hline
Instrumental systematic noise           & $3.0\cdot 10^{-1}$            \\
\hspace{1mm} First order phase            & $6.0\cdot 10^{-2}$         \\
\hspace{1mm} Second order amplitude-phase & $2.9\cdot 10^{-1}$           \\ \hline
Instrumental photon noise    & $1.8\cdot 10^{-2}$           \\
\hspace{1mm} Stellar null-floor leakage   & $1.0\cdot10^{-2}$            \\
\hspace{1mm} Polarisation angle           & $2.2\cdot10^{-3}$            \\
\hspace{1mm} Detector thermal background & \fis{$1.1\cdot10^{-3}$}  \\
\hspace{1mm} Detector dark current      & \fis{$1.5\cdot10^{-2}$}\\ \hline
SNR, one rotation (56 h)\tablefootmark{a}           & $4.5$           \\ \hline
\end{tabular}
\tablefoot{\tablefoottext{a}{Calculated by integrating over all wavelength bins for the specified integration time and can be compared to the value computed in Figure \ref{fig:methods_earth_twin_fluxes_and_snr} (SNR$\approx$9.7), where the same observing example was used, but without including instrumental noise terms.}}
\label{tab:4micron}
\end{table}

\begin{figure}[h]
    \centering
    \includegraphics[width=\linewidth]{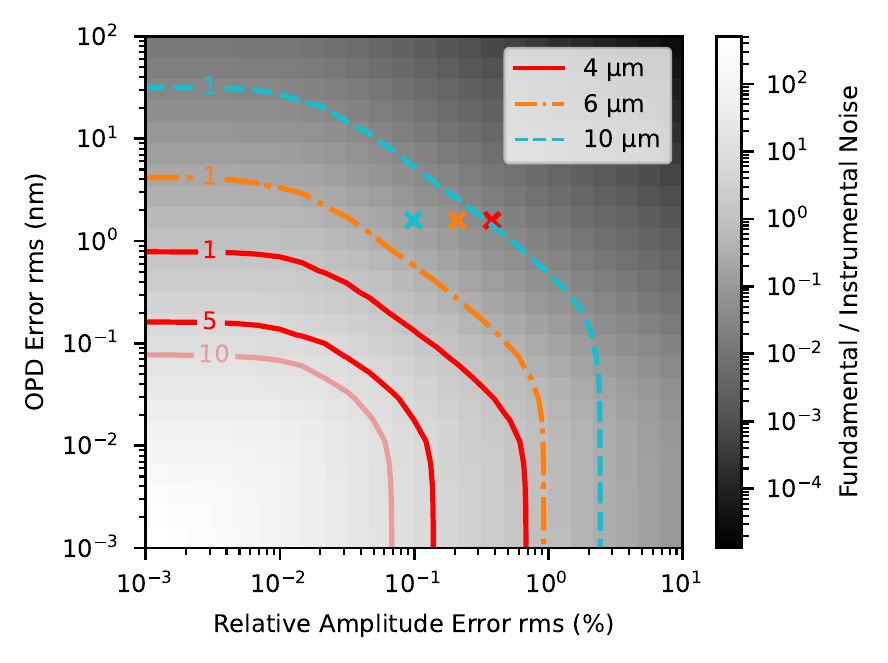}
    \caption{The ratio of fundamental noise to instrumental noise on a grid of amplitude and phase perturbations for the shortest-wavelength bin at $\lambda_b = \SI{4.1}{\micro\meter}$ with width $\Delta \lambda_b = \SI{0.2}{\micro\meter} $. The perturbations are assumed to be pink noise with the specified reference rms at \SI{10}{\micro\meter}, and the phase perturbation is converted to an optical path difference (OPD). The contours indicate where the fundamental noise dominates the systematic noise by a factor of one, five or ten respectively for the $\lambda_b = \SI{4.1}{\micro\meter}$ case (in red) and for comparison for the $\lambda_b = \SI{6.1}{\micro\meter}$ (in orange) and $\lambda_b = \SI{10.1}{\micro\meter}$ (in cyan) cases. The crosses indicate the reference levels specified by Table \ref{tab:noise}.}
    \label{fig:grid}
\end{figure}

To infer the acceptable systematic noise levels, we run the simulations over a grid of pink noise amplitude and phase perturbations with varying root-mean-squared (rms) values\footnote{We remind the reader that the use of phase chopping removes most of the coupling to other sources of perturbation.}. The rms is specified at a reference wavelength of \SI{10}{\micro\meter} and scaled according to Table \ref{tab:noise}. Phase errors are converted to optical path difference (OPD) errors to allow for a better comparison with literature. The ratio of instrumental to fundamental noise on this grid is shown in Figure \ref{fig:grid}. 

Firstly, the figure confirms that certain regions allow for a trade-off between the OPD error rms and the amplitude error rms, where increasing one allows for a reduction of the other.
Outside of these regions, there exist hard upper limits for the maximally allowed rms error for both amplitude and OPD. It will be vital for the design of the instrument to stay clear of these limits, since they imply that, if reached in one perturbation rms, any improvement to the other perturbation rms does not improve instrument performance. Additionally, we can reconfirm that the requirements are much less strict in the longer wavelength regime. \\
This grid can serve as a rough predictor for the maximally acceptable noise levels: acceptable and too high levels of noise are divided by the contour tracing a ratio of one between the fundamental and systematic noise. Within the presented framework, this produces sets of allowable noise levels shown in Table \ref{tab:results}. These values can only be seen as rough predictors, since we are not presenting a complete instrument model and additional noise sources not considered in the present analysis could further increase the noise requirements. \\

\begin{table}[h]
\centering
\caption{Approximate predictions for the maximum allowed perturbation rms to stay in the fundamental noise dominated case.}
\begin{tabular}{@{}ccc@{}}
\hline \hline
$\sigma_\mathrm{fund} = x \sigma_\mathrm{inst}$ & $x = 1$            & $x = 5$            \\
OPD error rms                & \multicolumn{2}{c}{\fis{Relative} amplitude error rms} \\
\fis{(nm)} & \multicolumn{2}{c}{\fis{(in \%)}} \\\hline
\textbf{0.75}                           & 0             & \ldots            \\
0.50                           & 0.02             & \ldots            \\
\textbf{0.16}                          & 0.08            & 0             \\
0.10                        & 0.13             & 0.02           \\
0.05                         & 0.24             & 0.05           \\
0.01                         & 0.62            & 0.12            \\
0                         & \textbf{0.67 }            & \textbf{0.14 }          \\\hline
\end{tabular}
\tablefoot{rms perturbations values are given relative to the reference wavelength of \SI{10}{\micro\meter}. Presented values are for an observation of an Earth-twin at 10 pc with LIFE configured in the baseline scenario in the $\lambda_b = \SI{4.1}{\micro\meter}$ wavelength bin, i.e. the given values follow the contours in Figure \ref{fig:grid}. The asymptotic upper limits are indicated by the bold text.}
\label{tab:results}
\end{table}

A similar analysis can be done for the instrumental photon noise sources. As representative terms for thermal background emission and white shot noise term we analyse the thermal emission as seen by the detector and a dark current term for the detector.

\begin{figure}[h]
    \centering
    \includegraphics[]{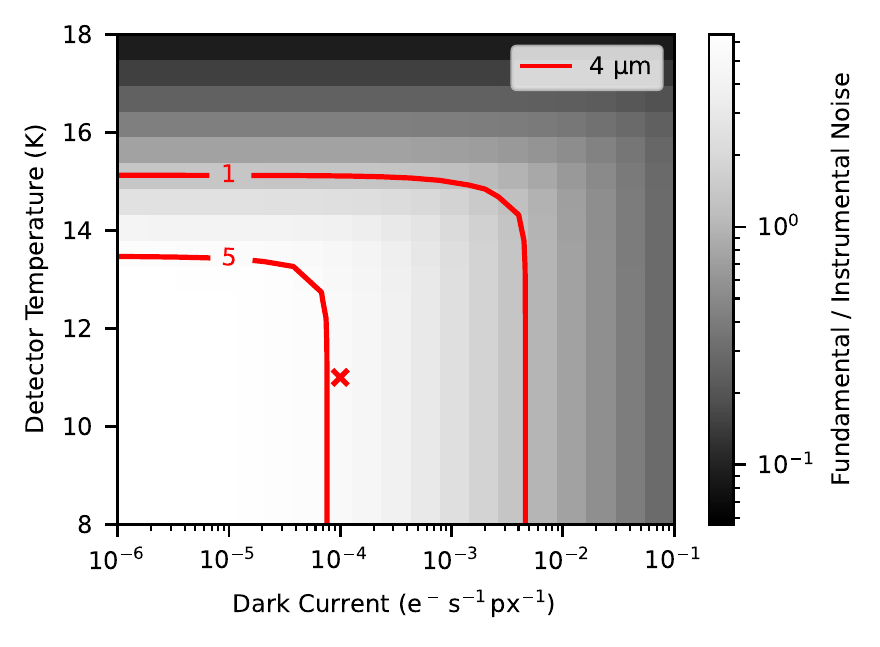}
    \caption{The ratio of fundamental noise to instrumental noise on a grid of the detector environment temperature and the detector dark current for the shortest-wavelength bin at $\lambda_b = \SI{4.1}{\micro\meter}$ with width $\Delta \lambda_b = \SI{0.2}{\micro\meter}$. The temperature induces black body thermal emission which is propagated directly into the detector. The dark current is assumed to be constant with temperature. The instrument is configured according to the reference state in Table \ref{tab:noise}. The contours indicate where the fundamental noise dominates the systematic noise by a factor of one or five. The cross indicates the reference levels of $T = \SI{11}{\kelvin}$ and $I_\mathrm{D} = 10^{-4} \, \, \mathrm{e^- \, px^{-1} \, s^{-1}}$ used in Figure \ref{fig:wavelength2m}.}
    \label{fig:thermal}
\end{figure}

Figure \ref{fig:thermal} shows an expected behaviour. Since both noise sources are independent of each other, there exists an absolute upper limit for the acceptable level of both terms (see Table \ref{tab:thermal}). 

We note that while meeting the requirement for the thermal background temperature of the detector does not appear too challenging, the requirements on the dark current seem to be at least one order of magnitude more stringent than what has been achieved so far for both Si:As IBC detector arrays and mid-infrared MCT devices \citep[e.g.,][]{rieke_mid-infrared_2015, cabrera_characterization_2019, roellig_mid-infrared_2020, Gaspar_2020}. The challenges related to MIR detector technology (and availability) has already been discussed in LIFE paper I \citep{paper1}. An interesting alternative approach could be kinetic inductance detectors (KIDs) \fis{or Transition Edge Sensor detectors (TES)} that may also be applicable for the LIFE wavelength range \citep[e.g.,][]{perido_extending_2020, Nagler_2021}. \\

\begin{table}[h]
\centering
\caption{Approximate predictions for the asymptotic upper limits in detector thermal background and detector dark current perturbation to stay in the fundamental noise dominated case.}
\begin{tabular}{@{}rcc@{}}
\hline \hline
$n_\mathrm{p} = x n_\mathrm{s}$ & $x = 1$ & $x = 5$ \\ \hline
Thermal background temperature (K)  & 15.1 & 13.4        \\
Detector dark current ($\mathrm{e^- \, s^{-1} \, px^{-1}}$)         & $4.6 \cdot 10^{-3}$ & $7.5 \cdot 10^{-5} $ \\ \hline
\end{tabular}
\tablefoot{\fis{Presented values are for an observation of an Earth-twin at 10 pc with LIFE configured in the baseline scenario in the $\lambda_b = \SI{4.1}{\micro\meter}$ wavelength bin, i.e. the given values correspond to the limits in Figure \ref{fig:thermal}.}}
\label{tab:thermal}
\end{table}

\begin{figure}[h]
    \centering
    \includegraphics[width=1.\linewidth]{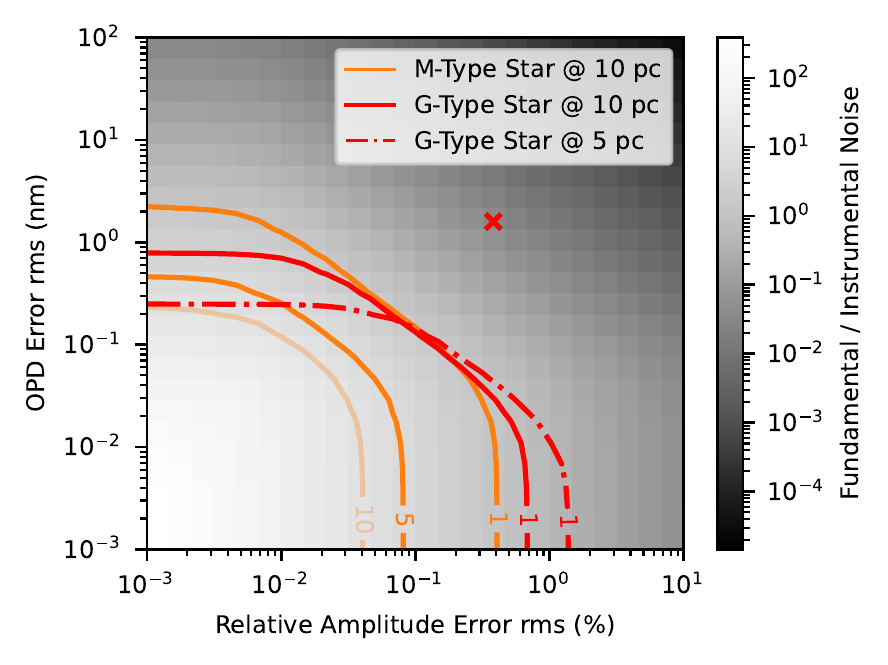}
    \caption{Impact of the target system configuration on the ratio of fundamental noise to instrumental noise on a grid of amplitude and phase perturbations for the shortest-wavelength bin at $\lambda_b = \SI{4.1}{\micro\meter}$ with width $\Delta \lambda_b = \SI{0.2}{\micro\meter} $. The system parameters assumed for the M-star target are given in the text. The contours indicate where the fundamental noise dominates the systematic noise by a factor of one, five or ten respectively. The cross indicates the $\SI{4}{\micro\meter}$ reference level specified by Table \ref{tab:noise}.}
    \label{fig:mstar}
\end{figure}

The acceptable perturbation levels will depend on the configuration of the observed system. First and foremost, the properties of the star and exo-zodical disk will influence both, the amount of fundamental noise received and how the amount of systematic noise scales with a given level of perturbation. Apart from this there are additional, more subtle effects that change the noise levels. Prominent examples are the scaling of the array baseline with the host star luminosity (see Section \ref{sec:SNR_calc_astro}), relative shifts in main emission wavelength between the astrophysical objects and scaling of absolute flux with the target distance. \\
A full study of all the aforementioned effects is vital, but out of the scope of this paper. Nevertheless, a simulation varying only the stellar type and the distance is performed below to confirm low impact on the acceptable noise levels and find an order-of-magnitude estimate for the relative change. \\
To simulate the observation around a different stellar type, an Earth-twin is placed around HD 232979\footnote{Effective temperature $T_\mathrm{eff} = \SI{4047}{\kelvin}$, effective radius $R_\mathrm{eff} = 0.57 \, \mathrm{R_\odot}$, luminosity $L = 0.079 \, \mathrm{L_\odot}$ \citep{gaia_collaboration_gaia_2016, gaia_collaboration_gaia_2018}, planet separation scaled to $a=0.31$ according to an planetary equilibrium temperature of $T_\mathrm{eq} = \SI{265}{\kelvin}$}, an M0.5V-type star at 10 pc distance. For this target as well as for an Earth-twin system at 5 pc distance, the ratio of fundamental to instrumental noise over amplitude and OPD perturbations is shown in Figure \ref{fig:mstar}.

It is apparent that the change in system configuration undertaken here changes the acceptable perturbation levels by less than one order of magnitude. Away from the amplitude and phase limits (towards the center of Figure \ref{fig:mstar}), the acceptable levels show almost no change. This can be taken as validation for the approach of this appendix to retrieve order-of-magnitude estimates using only a single representative Earth-twin target. However it must also be a reminder that this will propagate systematic biases into the sample of detectable planets, making an analysis of the full sample a top-priority for future work. \\
An additional interesting property found here is that decreasing the distance to the target increases the strictness of the OPD error requirement while decreasing the strictness of the amplitude error requirement. The same effect can be seen when moving to hotter host stars.

\subsection{Discussion}

In Table \ref{tab:results} and Table \ref{tab:thermal}, we present an estimate for the maximum allowed noise level to stay in the fundamental noise limited regime. We have cross-validated the simulations producing these results with \cite{Lay2004SystematicInterferometers}. While the maximum allowed noise levels can confidently be used as an order-of-magnitude approximation, we briefly discuss the most prominent factors that will influence these values as work on LIFE continues to progress. \\
First, in Figure \ref{fig:mstar} we demonstrate three different configurations of the target systems which do not produce significantly different requirements towards instrument perturbations. However, the \emph{LIFE} mission aims at observing a much more diverse sample than what is represented in Figure \ref{fig:mstar}. Therefore, in order to gain an understanding about the requirements to remain in the fundamental noise limited regime for every target, we will apply the presented simulation to the full synthetic planetary sample described in \cite{paper1} in future work. \\
Secondly, the presented analysis assumes a specific type of beam combination (double Bracewell). The \emph{LIFE} initiative is still in the process of evaluating other combiner techniques \citep{hansen2022large}. An adoption of a significantly different technique would certainly also change the requirements on maximum allowed perturbations and the types of perturbations that dominate the noise. \\
Thirdly, the assumptions about the perturbations themselves are not directly based on real data or a detailed instrument model. On the one hand, an approach more strongly based on experiments could inform a more detailed model for the shape of the perturbation spectra. As is shown in \cite{Defrere2010NullingMissions} the shape can significantly influence the requirements for maximal perturbations. On the other hand, it would enable the connection of requirements as presented in this report to a component level breakdown. This would enable an evaluation of the feasibility of reaching the required levels of perturbation. \\

Meeting these requirements for amplitude and especially OPD error described in Table \ref{tab:results} will be a considerable challenge. For comparison, currently planned lab experiments in the context of the \emph{LIFE} mission \citep{gheorghe_preparatory_2020} plan to achieve similar amplitude mismatch. The planned OPD error however is one order of magnitude larger compared to what is presented in this work. Based on these experiments it will be difficult to predict whether the requirements in Table \ref{tab:results} are achievable in the mid-term future. In turn, this raises the question if it is strictly required to operate the instrument in the fundamental noise limited regime over the whole wavelength range to accomplish the science goals proposed for the \emph{LIFE} mission. \\
Allowing for more OPD and amplitude perturbation will increase the instrumental noise at short wavelengths. However, these wavelength bands are routinely dominated by stellar geometric leakage (fundamental noise) and offer very little signal at least for temperate exoplanets (see Figure \ref{fig:methods_earth_twin_fluxes_and_snr}). Accordingly, for those targets, wavelengths below \SI{6}{\micro\meter} do not contribute to the bulk detection nor the atmospheric retrieval \citep[cf.][]{konrad2021large}. We therefore expect that allowing for appropriately larger perturbations will not significantly alter the results presented in this paper and in \cite{paper1}. \\

    \section{Implementation of maximum likelihood method}
    
    \begin{figure*}[]
        \centering
        \includegraphics[width=0.8\linewidth]{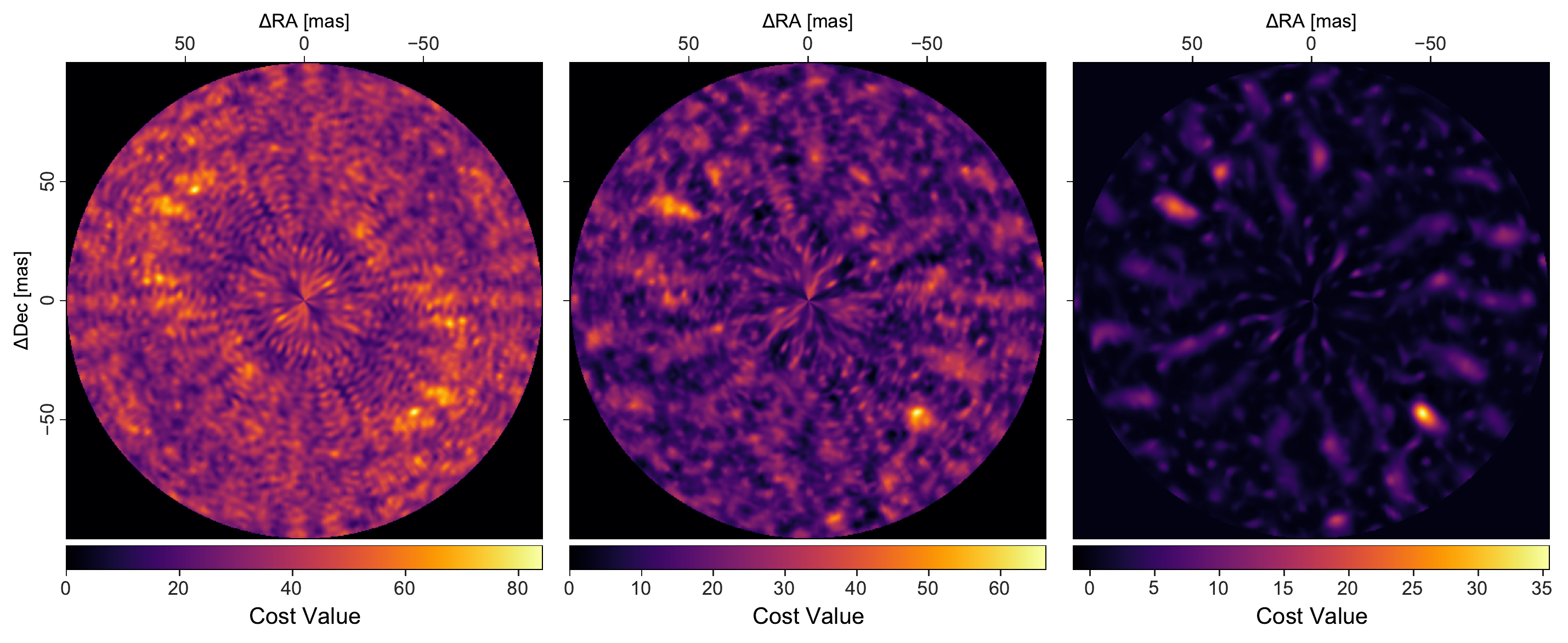}
        \caption{Different versions of the cost function $J$, which is calculated on a grid up to \SI{100}{mas} angular separation from the star. An artificial planet with a photon-based SNR$\sim5$ is located in the lower right at $\SI{67}{mas}$ from the center. \textit{Left}: $J'$ without positivity constraint on the flux. The image is symmetric and the contrast is low such that the planet position cannot be inferred. \textit{Middle}: $J''$ with a simple positivity constraint on the planet flux. The spatial degeneracy is removed and the most likely planet position in the lower right indicates the correct planet position. Two side lobes with similar cost value to the true planet position are visible. \textit{Right}: Cost function $J_{\mathrm{reg}}$ with positivity constraint on the flux and spectral regularization. The overall contrast is enhanced and the side lobes are suppressed compared with the unregularized cost function.}
        \label{fig:analysis_cost_evolution}
    \end{figure*} 
    We describe the implementation of the maximum likelihood method for the setting described in Sect.~\ref{sec:SNR_calc}, with symmetrical noise sources that contribute by shot noise only. The time-dependence of the transmission map originates only from the rotation of the array, such that $T(t, \lambda, \Vec{\theta}_\mathrm{p}) = T(\phi(t), \lambda, \Vec{\theta}_\mathrm{p})$. The ideal modulation profiles for all possible planet positions $\Vec{\theta}_\mathrm{p}$ are already contained in the transmission map shown earlier in Fig.~\ref{fig:pol_map_and_signal_profile} as horizontal lines. Additionally, the noise variance is time-independent: $\sigma^2(t,\lambda) = \sigma^2(\lambda)$ and can be estimated from the recorded data $A(t,\lambda)$.
    
    For simplicity, here $J$ labels the total cost function including the regularization term. To enforce smoothness on the spectrum, the second order derivative $m=2$ is used.
    The discretized cost function is given by
    \begin{align}\label{eq:cost_function_disc}
        \begin{split}
        J(\Vec{\theta}, \Vec{F}) =
        \sum_{\lambda} \Bigg[
        &\sum_t 
        - 2 F_{\lambda} \sum_t \dfrac{A_{t,\lambda}\, T_{t, \lambda} }{\sigma^2_{\lambda}}
        + F_{\lambda}^2 \sum_t \dfrac{T^2_{t, \lambda} }{\sigma^2_{\lambda}} \\
        & + \mu \, \left( \dfrac{F_{\lambda+1} - 2 F_{\lambda} + F_{\lambda-1}}{(\Delta \lambda)
^2} \right)^2 \Bigg]
        \end{split} 
    \end{align} 
    where the dependence of $A$ and $T$ on the considered position $\Vec{\theta}_\mathrm{p}$ is omitted in the notation on the right hand side. To switch to matrix notation, the following parameters are defined:
    \begin{align}
        \Vec{c} &:= \{c_\lambda\}\,; \qquad \quad \mathrm{where} \; c_\lambda = \sum_t A_{t,\lambda}\,  T_{t, \lambda} / \sigma^2_{\lambda}\\
        \mathbf{B} &:= \operatorname{diag} \{ B_\lambda \}\,;  \quad \mathrm{where} \;  B_\lambda = \sum_t T^2_{t, \lambda} / \sigma^2_{\lambda}
    \end{align}
    
    \begin{equation}
    \mathbf{D} = 
    \begin{pmatrix}
    -1 & 1 & \cdots & \cdots &\cdots & \cdots \\
    1 & -2 & 1 & \cdots &\cdots & \cdots \\
    \vdots & 1 & -2 & 1 & \cdots & \cdots \\
    \vdots & \vdots & \ddots & \ddots & \ddots & \vdots\\
    \vdots & \vdots & \vdots & \ddots & -2 & 1\\
    \vdots & \vdots & \vdots & \vdots & 1 & -1

    \end{pmatrix}
    \end{equation}
    where $\Vec{c}$ corresponds to the unregularized flux estimate vector, and $\mathbf{D}$ is the second order derivation matrix, which is symmetric. Equation~\eqref{eq:cost_function_disc} can then be rewritten in matrix notation as
    \begin{align}\label{eq:cost_func_vec}
        J(\Vec{\theta}, \Vec{F}) &=
            -2 \Vec{F} \cdot \Vec{c} + \Vec{F}^\intercal \mathbf{B} \Vec{F} + 
            \mu \, \lVert \mathbf{D} \Vec{F} \rVert^2,
    \end{align}
    The condition for minimizing the cost function with respect to the flux, Eq.~\eqref{eq:cost_opt_criterion}, then
    directly returns the optimized flux vector as 
    \begin{align}
        \Vec{\hat{F}} = (\mathbf{B} + \mu\,\mathbf{D}^2)^{-1}  \Vec{c} \label{eq:_opt_flux_vec}
    \end{align}   
    which now also includes the regularization enforced by $\mathbf{D}$.
    If the optimized flux is inserted back into Eq.~\eqref{eq:cost_func_vec}, we get the cost function $J'(\Vec{\theta}) = J(\Vec{\theta}, \Vec{\hat{F}})$, which is optimized with respect to the flux: 
    \begin{align}\label{eq:cost_func_vec2}
        J'(\Vec{\theta}, \Vec{\hat{F}})
            &= - \Vec{c}^\intercal  (\mathbf{B} + \mu\,\mathbf{D}^2)^{-1} \Vec{c} = - \Vec{\hat{F}} \cdot \Vec{c} 
    \end{align}   
    where only \Vec{c} depends on the recorded data.
    As with the unregularized cost function in Sect.~\ref{sec:MLM}, a positivity constraint can be applied on the regularized estimated flux:
    \begin{align}
        J''(\Vec{\theta}) &= J[\Vec{\theta}, \Vec{\hat{F}}_\mathrm{pos}(\Vec{\theta})]
        = - \Vec{\hat{F}}_\mathrm{pos} \cdot \Vec{c} 
    \end{align}
    The computation of the likelihood maps can be performed on any grid of possible planet positions and does not necessarily have to cover the whole field-of-view of the telescope.
    
    In Fig.~\ref{fig:analysis_cost_evolution} we show a concrete example for how the different cost functions behave.

    \section{Local zodiacal emission} \label{app:localzodi}

\begin{figure}[ht]
    \centering
    \includegraphics[width=0.75\linewidth]{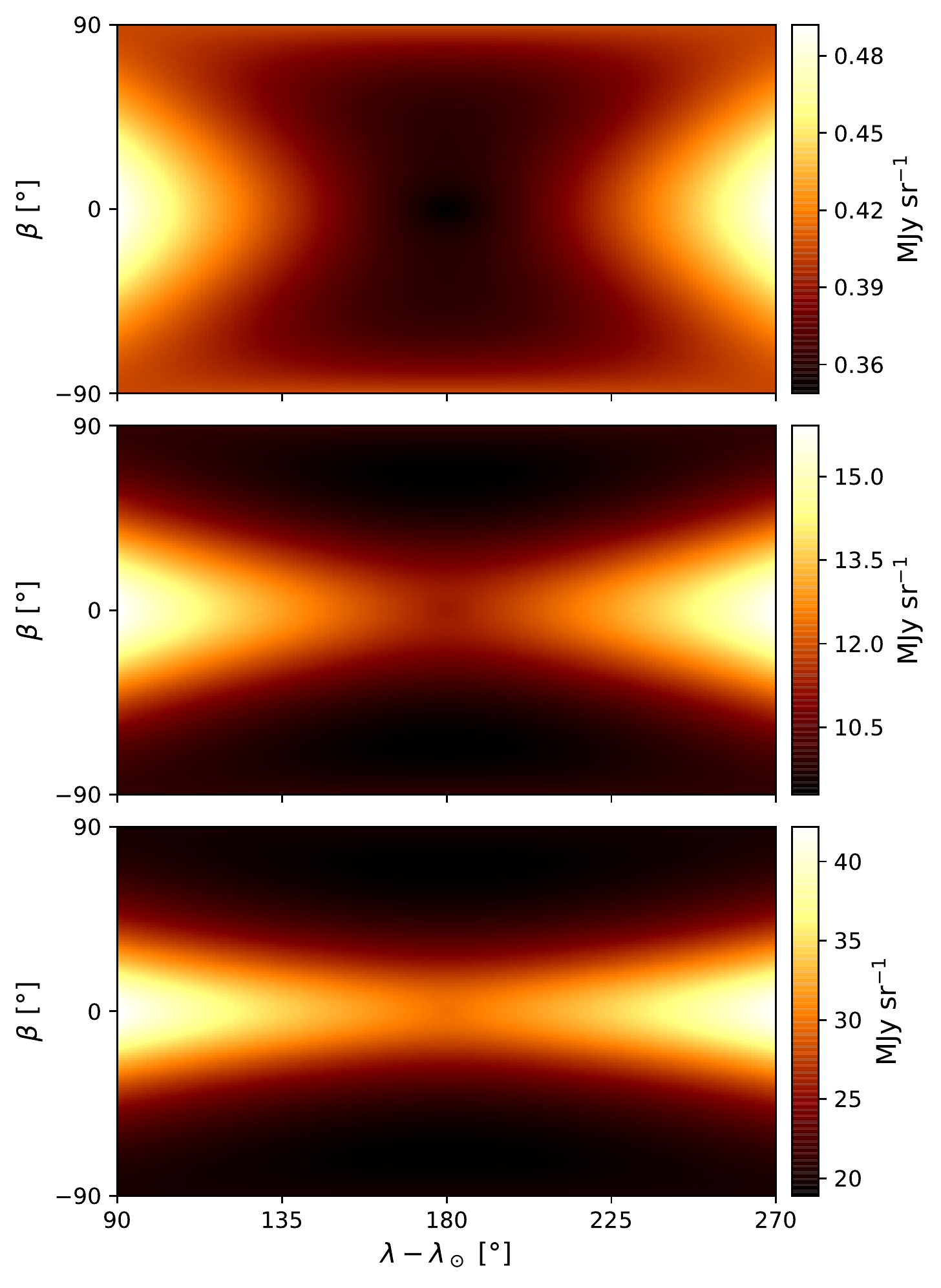}
    \caption{Surface brightness distribution of local zodiacal dust cloud for wavelengths of \SI{5}{\micro\meter}, \SI{10}{\micro\meter}, and \SI{20}{\micro\meter} (from top to bottom). The flux is given in MJy\,sr$^{-1}$ to allow for a comparison with the \textsc{DarwinSim} technical report \citep{Hartog2005TheSimulator} and other models for the local zodiacal dust, e.g.,  the one used for sensitivity predictions for MIRI instrument onboard the \emph{JWST} \citep{Glasse2015Sensitivity}. We note the different  colorscale for each wavelength.}
\label{fig:lz_dist}
\end{figure}

Figure \ref{fig:methods_earth_twin_fluxes_and_snr} indicates that the local-zodiacal thermal emission is the dominant astrophysical noise source in \emph{LIFE's} long-wavelength regime. Figure \ref{fig:lz_dist} depicts a visual representation this emission according to Formula \eqref{eq:LZdarwinsim}. It shows that with respect to local-zodi leakage it is generally favourable to observe targets at high ecliptic latitudes and longitudes close to the anti-sun direction.

\end{document}